\documentclass[fleqn, twocolumn, amsmath,amssymb,superscriptaddress]{revtex4-1}
\usepackage{epsfig}
\usepackage{amsmath}
\usepackage{graphicx}
\usepackage{color}
\usepackage{amsfonts,amssymb}
\usepackage[thinlines]{easytable}
\usepackage[english]{babel}
\usepackage{hyperref}
\usepackage{lipsum}
\usepackage{graphics}
\usepackage{array}
\usepackage{tabu}
\usepackage{dsfont}

\newcommand{\ct}{\cite}

\newcommand{\be}{\begin{equation}}
\newcommand{\ee}{\end{equation}}
\newcommand{\ba}{\begin{eqnarray}}
\newcommand{\ea}{\end{eqnarray}}

\newcommand{\non}{\nonumber}
\newcommand{\bra}[1]{\langle #1|}
\newcommand{\ket}[1]{|#1\rangle}

\begin{document}

\title{Enhanced precision bound of low-temperature quantum thermometry via dynamical control}

\author{Victor Mukherjee}
\email{mukherjeev@iiserbpr.ac.in}
\affiliation{International Center of Quantum Artificial Intelligence for Science and Technology (QuArtist) \\ and Department of Physics, Shanghai University, 200444 Shanghai, China}
\affiliation{Department of Chemical and Biological Physics, Weizmann Institute of Science, Rehovot 7610001, Israel}
\affiliation{Department of Physical Sciences, IISER Berhampur, Berhampur 760010, India}

\author{Analia Zwick}
\affiliation{Departamento de  F\'isica M\'edica, Centro At\'omico Bariloche, CNEA, CONICET, 8400 S. C. de Bariloche, Argentina}

\author{Arnab Ghosh}
\affiliation{International Center of Quantum Artificial Intelligence for Science and Technology (QuArtist) \\ and Department of Physics, Shanghai University, 200444 Shanghai, China}
\affiliation{Department of Chemical and Biological Physics, Weizmann Institute of Science, Rehovot 7610001, Israel}
\affiliation{Department of Chemistry, IIT Kanpur, Kanpur 208016, India}

\author{Xi Chen}
\affiliation{International Center of Quantum Artificial Intelligence for Science and Technology (QuArtist) \\ and Department of Physics, Shanghai University, 200444 Shanghai, China}
\affiliation{Department of Physical Chemistry, University of the Basque Country UPV/EHU, Apartado 644, 48080 Bilbao, Spain}

\author{Gershon Kurizki}
\affiliation{Department of Chemical and Biological Physics, Weizmann Institute of Science, Rehovot 7610001, Israel}

\begin{abstract}
High-precision low-temperature thermometry is a challenge for experimental quantum physics and quantum sensing. Here we consider a thermometer modelled by a dynamically-controlled
multilevel quantum probe in
contact with a bath. Dynamical control in the form of periodic modulation of the energy-level spacings of the quantum probe
can  dramatically increase the maximum accuracy bound of low-temperatures estimation, 
by maximizing the relevant quantum Fisher information. 
As opposed to the
diverging relative error bound at low temperatures in conventional quantum thermometry, periodic modulation of the probe allows for  low-temperature thermometry
with temperature-independent relative error bound.
The proposed approach may find diverse applications related to precise probing of the temperature of many-body quantum systems in condensed matter and ultracold gases,
as well as in different branches of quantum metrology beyond thermometry, for example in 
precise probing of different Hamiltonian parameters in many-body quantum critical systems.

\end{abstract}

\maketitle

\section*{Introduction}
Precise probing of quantum systems is one of the keys to progress in diverse quantum technologies, including quantum metrology \cite{giovannetti11, braun18quantum, kurizki15quantum}, quantum information 
processing (QIP) \cite{hauke16} and 
quantum many-body manipulations \cite{strobel14}. The maximum amount of information obtained on a 
parameter of a quantum system is quantified by the quantum Fisher information (QFI), which depends on the extent to which the state of the system changes for an infinitesimal change in the estimated parameter 
 \cite{paris09, caves94, correa15, zwick16, pasquale16}. Ways to increase the QFI, thereby increasing the precision bound of parameter estimation, are therefore recognized to be
of immense importance \cite{gefen17, campbell18precision}. Recent works have studied QFI for demonstrating the criticality of environmental (bath) information \cite{zwick16pra}, QFI
enhancement in the presence of strong coupling \cite{correa16} or by dynamically-controlled quantum probes
\cite{zwick16} and the application of quantum thermal machines to
quantum thermometry \cite{brunner17}. 

Here we propose the synthesis of two concepts: quantum 
thermometry \cite{matteo12, correa15, correa16, brunner17, campbell18precision, potts19fundamental, hovhannisyan18measuring, kiilerich18dynamical, tuoriniemi16physics,
mehboudi19thermometry, pasquale19quantum} and temporally-periodic dynamical control
that has been originally developed for decoherence suppression 
in QIP \cite{agarwal99, agarwal01, lidar04,  zwick14, dieter16}. We show that such control can 
strongly increase the 
 QFI that determines the precision-bound of temperature measurement, particularly at temperatures approaching absolute zero. Accordingly, such control
 may boost the ultralow-temperature precision bound of diverse thermometers, e.g. those based on Coulomb 
blockade \cite{meschke11comparison},
 metal - insulator - superconductor junctions \cite{raisanen18normal}, kinetic inductance \cite{giazotto08ultrasensitive}
and novel hybrid superconductors \cite{giazotto15ferromagnetic}. Alternatively, dynamical control may allow theese thermometers 
 to accurately estimate a broad range of temperatures.

\section*{Results}
\begin{figure}[t]
\begin{center}
\includegraphics[height=1.5in,width=3.2in]{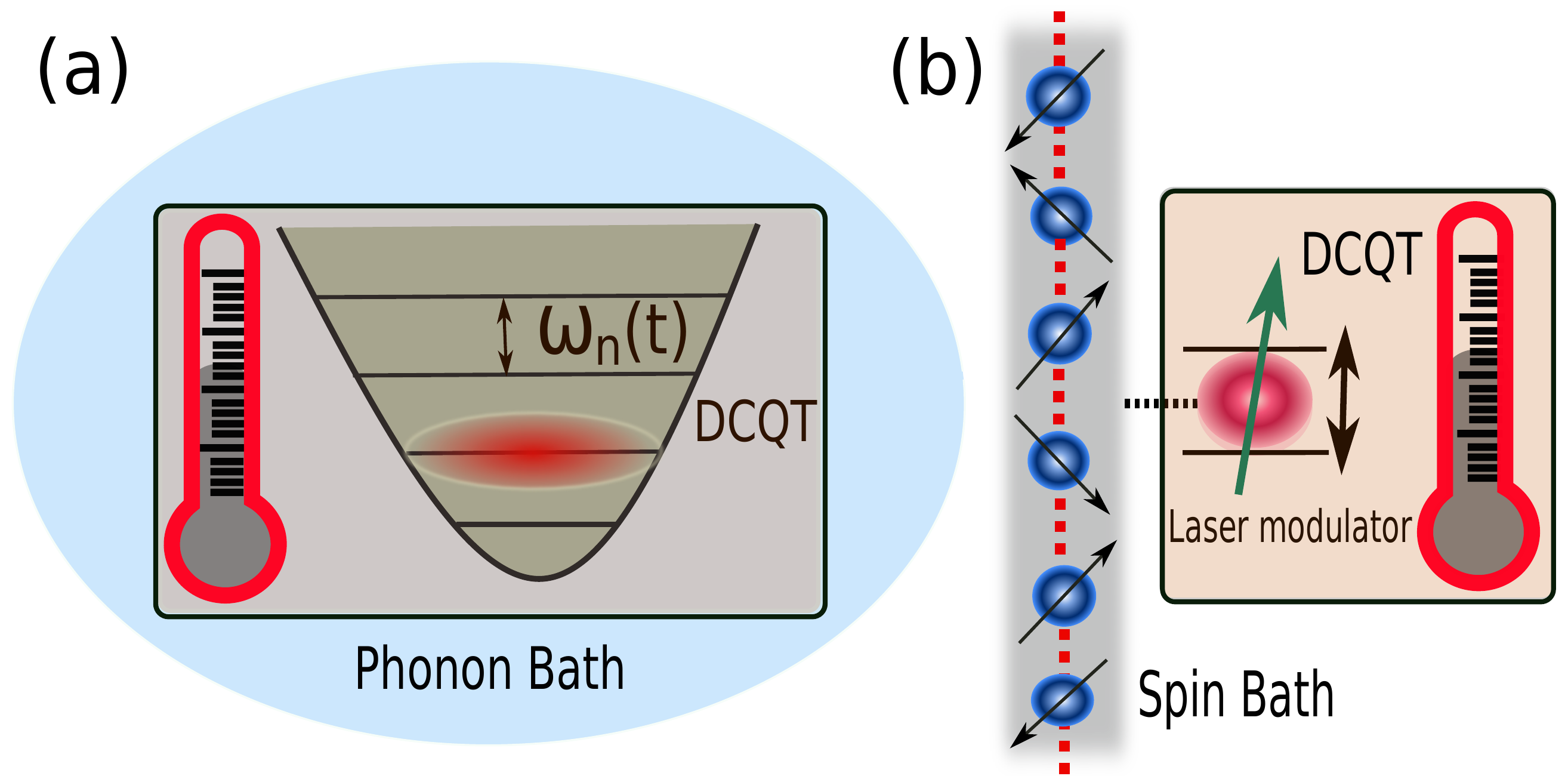}
\end{center}
\caption {{\bf Schematic realization of a quantum thermometer:} (a) Schematic realization of a dynamically controlled quantum thermometer by a motional wavepacket of
a state of a cavity, or an ion/atom trapped in an optical lattice potential. The frequency of the trap is periodically varied (black arrow)
by modulating the length of the cavity, or the amplitudes of the lasers forming 
the trap. The wavepacket is in contact with a  phonon bath, and relaxes to the corresponding thermal steady state  at long times. 
(b) Same, for a spin-chain bath coupled to a two-level system whose resonant frequency is periodically modulated.}
\label{fig2l}
\end{figure}

\noindent {\bf Model:} An example of our dynamically controlled quantum thermometer (DCQT) is a quantum wavepacket trapped in a  potential
and subjected to periodic modulation, while it is
 immersed in a thermal bath (Fig. \ref{fig2l}a). Measurements of the wavepacket after it has reached a steady state provide information about the bath temperature. 
Another example of a DCQT is the internal state of either a two-level or a multilevel system coupled to a spin-chain bath \cite{ghosh11, ghosh12}. 
The level separation is periodically modulated by a control field (e.g. a field-induced alternating current Stark shift)
and its level populations are finally read out by laser-induced fluorescence
in the optical range (Fig. \ref{fig2l}b). Many of the advanced thermometers \cite{correa16, brunner17, matteo12, meschke11comparison, raisanen18normal, giazotto08ultrasensitive, giazotto15ferromagnetic} may be adapted to such dynamically controlled operation.

The proposed DCQT is described by a Hamiltonian $H(t)$, subjected to a 
modulation with time period $\tau = 2\pi/\Delta$:
\ba
\hat{H}(t + \tau) &=& \hat{H}(t) = \sum_k \hat{H}_k(t);\non\\
\hat{H}_k(t+\tau) &=& \hat{H}_k(t).
\label{hamilN}
\ea
The DCQT  is coupled to a bath through the interaction Hamiltonian of the form $\hat{H}_I = \sum_{k} \hat{H}_{Ik} = \sum_k \hat{S}_k\otimes \hat{B}_k$.
Here $\hat{S}_k$ and $\hat{B}_k$ are, respectively, the $k$-th mode DCQT and bath operators.
In cases where the  $k$-th bath mode
acts as a Markovian environment characterized by a mode-dependent unknown temperature $T_k$, the corresponding DCQT mode thermalizes to the temperature $T_k$ at long times, thus enabling us 
to perform 
mode-dependent thermometry.
Alternatively, if the bath is thermal with
an unknown temperature $T$, a single-mode DCQT suffices for bath temperature estimation.

\vspace{0.2cm}
\noindent {\bf Analysis:} For simplicity,  we restrict our analysis below to a thermal bath with temperature $T$, probed by an  $\mathcal{N}$-level DCQT that is described by the Hamiltonian 
$\hat{H}(t) = \hbar\sum_{n = 0}^{\mathcal{N}} \omega_n(t)\ket{n}\bra{n}$ with energy spectrum  $\omega_{n}(t) = n\omega(t)$ for
real positive 
$\omega(t)$, and an interaction Hamiltonian $\hat{H}_I = \hat{S}\otimes\hat{B}$. Here $n$ denotes the level index, and we assume $\hat{S}$ to be a system operator belonging to a Lie algebra.  
For a two-level system, $\hat{S} \equiv \hat{\sigma}_x$, the $x$ Pauli operator, while for a 
harmonic oscillator, $\hat{S} \equiv \left(\hat{a} + \hat{a}^{\dagger}\right)$, where $\hat{a}$, $\hat{a}^{\dagger}$ denote the annihilation and creation operators, respectively.
For a single-mode DCQT, we consider a generic, periodic, diagonal (frequency) modulation:
\ba
\omega(t) &=& \sum_{m^{\prime} = 0}^{\infty} \Big(s(m^{\prime}) \sin(m^{\prime}t \Delta) \non\\
&+& c(m^{\prime}) \cos(m^{\prime}t \Delta) \Big),
\label{modgen}
\ea
where  $s(m^{\prime}), c(m^{\prime})$ are adjustable real constants, corresponding to the $m^{\prime}$-th frequency harmonic. One can extend this analysis to a multi-mode
DCQT probing a multi-mode bath (see Methods).

Here we focus on controls with $\tau$ much smaller than the thermalization time of the system, such that one can
adopt the secular 
approximation, thereby averaging out all rapidly
oscillating terms and arriving at a time $t \gg \tau$ at the steady state 
$\hat{\rho}(t \to \infty) = \sum_{n = 0}^{\mathcal{N}} \varrho_{n} \ket{n}\bra{n}$ \cite{klimovsky13, alicki14, klimovsky15}.
The level populations  $\varrho_n$ are functions of the bath temperature $T$, the bath response (correlation) functions
$G(\omega_m)$ (see Methods) \cite{breuer02, yan18distinguishing} at the $m$-th sideband frequency $\omega_m = \omega_0 + m\Delta$,
and the  modulation-dependent $m$-th sideband weight $P_m \geq 0$ (see Methods and Apps. \ref{appA} - \ref{appE}). 
Here the sidebands $m= 0, \pm 1, \pm 2, \ldots$ arise due to the
periodic modulation, the mean frequency
 $\omega_0  = \left(\int^{\tau}_0 \omega(t) dt\right)/\tau$, and $P_m$'s satisfy the constraint $\sum_m P_m = 1$ \cite{klimovsky13, alicki14, klimovsky15}. 
As can
 be expected in experiments, we impose an upper bound on the available resources for dynamical control, by considering only 
frequency modulations such that ${\rm max}\left[m^{\prime}\right]\Delta < \omega_0$.  Unless otherwise stated, we take
$\hbar$ and $k_{B}$ to be unity, and assume 
the bath response obeys the standard Kubo-Martin-Schwinger condition
$G(-\omega)/G(\omega) = {\rm exp}\left(-\omega/T \right)$ \cite{breuer02}.

We may infer the bath temperature $T$ from measurements of $\varrho_n$ of the thermometer: e.g., through measurement of the average phonon occupation number 
of a trapped-wavepacket probe (Fig. \ref{fig2l}a) \cite{raymer04experimental}, or the fluorescence of a
two-level system probe (Fig. \ref{fig2l}b). 
We can assess the effectiveness of our measuring scheme from the QFI 
as a function of the system and bath parameters.
The relative
 error $\delta T/T$ is bounded by the minimal achievable error $\xi$, dictated by the Cramer-Rao bound for optimal positive-operator valued measure, which in this case are the level-population measurements. It
 obeys the relation \ct{paris09, brunelli11, zwick16}
\ba
\frac{\delta T}{T} \geq \xi = \frac{1}{T\sqrt{\mathcal{M} \mathcal{H}}}.
\label{errb}
\ea
Here $\mathcal{M}$ denotes the number of measurements, and $\mathcal{H}  = -2\lim_{\epsilon \to 0}\partial^2 F(\hat{\rho}(T,t), \hat{\rho}(T+\epsilon,t))/\partial \epsilon^2 = \sum_{n = 0}^{\mathcal{N}} |\partial\varrho_{n}/\partial T|^2/\varrho_{n}$ is the 
QFI at temperature $T$ \cite{caves94, pezze09, Kacprowicz10, correa15},
 $F(\hat{\rho}_1, \hat{\rho}_2) = {\rm Tr}\left[\sqrt{\sqrt{\hat{\rho}_1}\hat{\rho}_2\sqrt{\hat{\rho}_1}}\right]$ being the fidelity between $\hat{\rho}_1$ and $\hat{\rho}_2$.

The dynamics of the DCQT, and consequently also the resultant steady state, depends crucially on the factors 
$P_{m} G(\omega_m)$; in general, a large $P_{m} G(\omega_m)$ is beneficial for estimating temperatures $T \sim \omega_m$ (see below, Methods and Apps \ref{appA} - \ref{appE} Notes 1 - 5) \cite{klimovsky15}.
Although we cannot control the bath correlation functions (spectral response) $G(\omega_m)$, nevertheless
for a given $G(\omega_m)$, 
we can tailor the thermometry 
QFI to our advantage by judicious choices of the periodic control fields, and hence of the corresponding $P_m$'s and $\omega_m$'s.
For example, a
versatile thermometer which can measure a 
wide range 
of temperatures with high accuracy bound would require a modulation that corresponds to 
large QFI over a broad range of temperatures, in order to yield $\xi \ll 1$ for finite $\mathcal{M}$. 
The above scenario can be realized using  modulations which give rise to multiple non-negligible $P_m$'s, or equivalently, large $P_m G(\omega_m)$ over a broad range of frequencies.
In particular, it is exemplified below  using a sinusoidal modulation characterized by a single, or few frequencies.   On the other hand,
a different modulation would enable  the  same 
thermometer to measure temperatures with higher accuracy bound (or, equivalently, larger QFI), 
but at the
expense of the changing of the temperature range 
 over which the DCQT can measure accurately. This can for example be realized
using periodic $\pi$-pulses, which give rise to only two sidebands, viz, non-negligible $P_{\pm 1}$ \cite{klimovsky13}.
 
Therefore, in contrast to thermometry in the absence of any control, DCQT 
 gives us the possibility of tuning the range and accuracy bound of  measurable temperatures. This necessitates an appropriate choice of control parameters 
 ($\Delta, m^{\prime}, s(m^{\prime}), c(m^{\prime})$) in Eq. \eqref{modgen} depending on the temperatures of interest.

\begin{figure}[t]
\begin{center}
\includegraphics[width = \columnwidth]{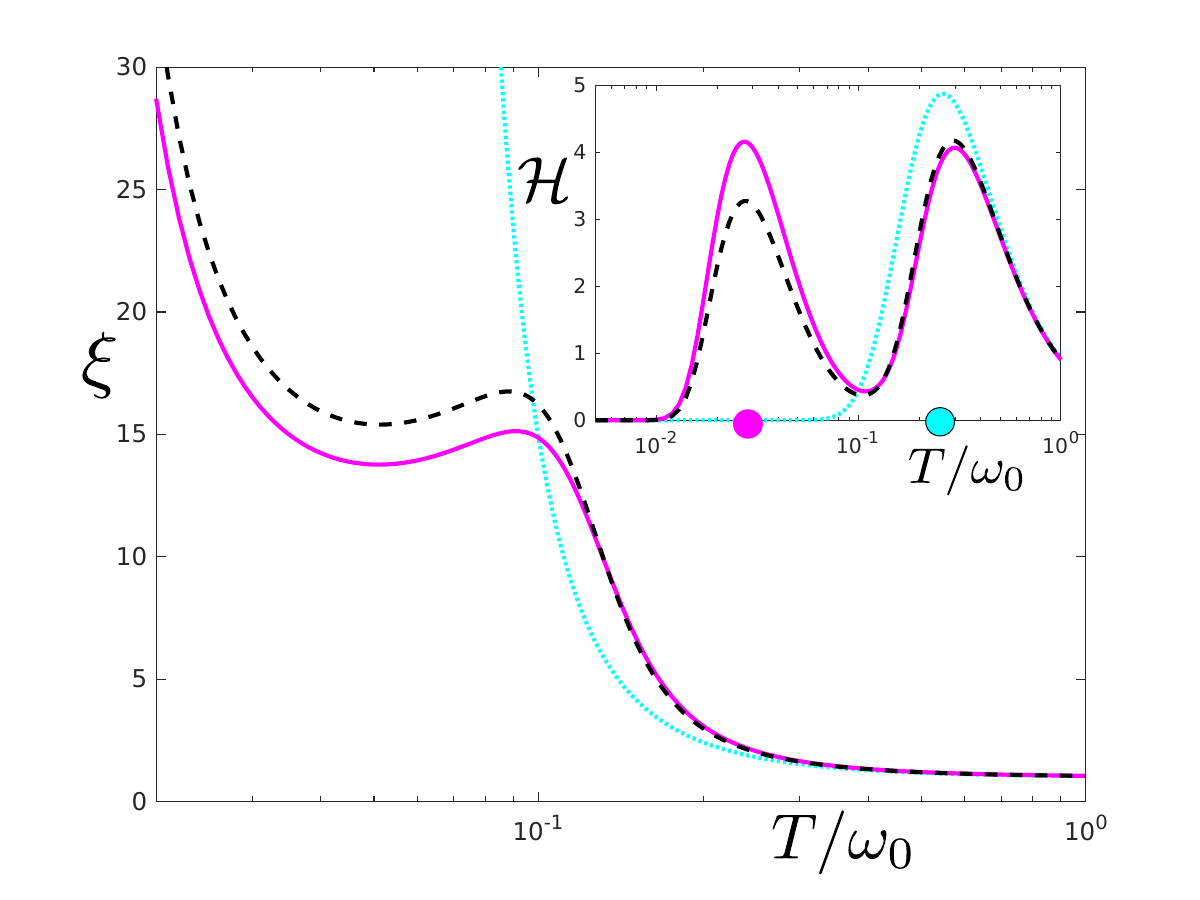}
\end{center}
\caption {{\bf Relative error bound and quantum Fisher information:} Relative error bound $\xi$ for estimation of bath temperature $T$ (in units of $\omega_0$)  by a harmonic
oscillator dynamically controlled quantum thermometer (DCQT) under sinusoidal modulation for the following bath spectra: nearly flat bath spectrum (magenta solid curve),  
sub-Ohmic bath spectrum with $s = 0.1, \omega_c = 100$ (black dashed curve) and the same spectra in the absence of control (turquoise dotted curve). For low temperatures in
the absence of control, $\xi \to \infty$, thus making it impossible to measure low temperatures as shown.
In contrast, our dynamical control scheme 
reduces $\xi$ to finite values  at low temperatures, for nearly flat, as well as sub-Ohmic bath spectra, thus showing the advantage of DCQT. Inset: Quantum Fisher Information ($\mathcal{H}$) as a 
function of temperature $T$  (in units of $\omega_0$) for DCQT under sinusoidal modulation and in the absence of control, for the bath spectra in the main figure (same curve colors). 
 DCQT increases the quantum Fisher information significantly at lower temperatures, giving rise to a peak at $T \approx T_{-1}$ (magenta dot),  in
addition to the peak at $T \approx T_0$ (turquoise dot). 
Here  the modulation amplitude $\mu = 0.2, \omega_0 = 1, \Delta = 0.9$ (see Eq. \eqref{sinmod}), 
$m = 0,\pm 1,\pm 2,\pm 3$ and the number of measurements $\mathcal{M} = 1$. The  thermalization time $\sim \gamma^{-1}$ is assumed to be long enough such that the secular 
approximation is valid.}
\label{figqfi}
\end{figure}

\vspace{0.2cm}
\noindent {\bf Harmonic-oscillator DCQT for sub-Ohmic baths:} As a generic example of DCQT, we consider the probe to be  a periodically modulated harmonic oscillator.
The interaction Hamiltonian is $\hat{H}_{I} = \left(\hat{a} + \hat{a}^{\dagger} \right)\otimes \hat{B}$,
where $\hat{a}$ ($\hat{a}^{\dagger}$) denotes the annihilation (creation) operator of the DCQT, and $\hat{B}$ is a bath operator.

We first consider the thermometry of a broad class of  bath spectral-response functions \cite{hovhannisyan18measuring,  milotti02onebyf} 
\ba
G(\omega) = \gamma \frac{\omega^s}{\omega_c^{s-1}}e^{-\omega/\omega_c}~~~\text{for $\omega \geq 0$},
\label{appspec}
\ea
under the Kubo-Martin-Schwinger  condition. Here $\gamma$ is a positive constant determining the system-bath coupling strength. Since we focus on the weak-coupling limit, such that the 
thermalization time $\sim \gamma^{-1} \gg \tau, \omega_{m}^{-1}$, the secular approximation is valid \cite{breuer02}. Expression \eqref{appspec} yields sub-Ohmic, Ohmic and super-Ohmic bath spectra for 
$s < 1$, $s = 1$ and $s > 1$ 
respectively. 

To show the full advantage of our control scheme, we start by focusing on a sub-Ohmic bath spectrum, since it has a non-zero $G(\omega)$ infinitesimally close to $\omega = 0$, even 
though $G(\omega)$ 
vanishes at $\omega = 0$  (Cf. Eq. \eqref{appspec}),
 which is ideal for low-temperature 
thermometry \cite{potts19fundamental}. To simplify the treatment, we take the limit of small $\gamma$ in Eq. \eqref{appspec} to
ensure weak coupling, $s\to 0$ to generate small $\omega_{\rm min}$, and $\omega_c \to \infty$ to realize high cutoff frequency, and arrive at
the nearly-flat bath spectrum (NFBS) 
\ba
G(\omega) \approx G_0 \approx \gamma s^s \omega_c \exp (-s)
\label{gnfbs}
\ea
for
$\omega_{\rm min} \lesssim \omega \ll \omega_c$, with
$\omega_{\rm min}  \approx s \omega_c$ (see Apps \ref{appD} and \ref{appE}). Hence, $\omega_{\rm min}$ can be as small as $s \omega_c$, thus enabling us to probe 
the bath very close to the absolute zero (see below). 
The comparison of our results with those obtained for a sub-Ohmic bath spectrum with finite $s$ and $\omega_c$ is shown in Figs. \ref{figqfi} and \ref{figerrgw}. The 
extension of our results to the case of a generic bath spectrum is
presented below.

Below we choose the  
control scheme Eq. \eqref{modgen} to have $c(m^{\prime}) = \omega_0 \delta_{m^{\prime},0}$ and 
$s(m^{\prime}) = \mu \Delta \delta_{m^{\prime},1}$, so that $\omega(t)$ is varied sinusoidally:
\ba
\omega(t) &=& \omega_0 + \mu\Delta \sin\left(t \Delta \right).
\label{sinmod}
\ea

In the limit of weak-modulation ($0 \leq \mu \ll 1$), only the sidebands $m = 0$ and $m = \pm 1$ have significant weights $P_{m}$ in the harmonic (Floquet) expansion of the system response \cite{klimovsky13}: 
 the $m$th sideband weight $P_m$, corresponding to the Floquet (harmonic) frequency 
$\omega_{m} = \omega_0 + m\Delta$, falls off rapidly with increasing $|m|$.
The QFI can then be written as (see App. \ref{appD})
\ba
\mathcal{H} &=& \mathcal{H}_{-1} + \mathcal{H}_{0} + \mathcal{H}_{1};\non\\
\mathcal{H}_{\pm 1} &=& \frac{P_{\pm 1}e^{\frac{\omega_{\pm 1}}{T}} \omega_{\pm 1}^2}{\left(-1+e^{\frac{\omega_{\pm 1}}{T}}\right)^2 T^4}, 
\label{chi3}
\ea
Here $\mathcal{H}_{\pm 1}$ are the contributions to the QFI arising from to the $m = \pm 1$ sidebands with $P_{\pm 1} \simeq \mu^2/4$,  
and $\mathcal{H}_{0}$ includes the $m = 0$ contribution.  

The advantages of DCQT are then revealed for a control scheme with $\Delta$ chosen such that 
\ba
\omega_{-1} \sim T \ll \omega_0, \omega_1,
\label{Treg}
\ea
so that 
\ba
\varrho_n = \frac{1}{N_{\rm avg} + 1}\left(\frac{N_{\rm avg}}{N_{\rm avg} + 1} \right)^n
\label{rhoneff}
\ea
with the average occupation number given by
\ba
N_{\rm avg} &=& N_{\rm eff} \approx \frac{1}{\frac{1}{P_1}e^{\omega_{-1}/T} - 1}
\label{rhossmu}
\ea
 The QFI is then predominantly associated with the $\mathcal{H}_{-1}$ term, with the 
maximum located at the temperature
\ba
T \approx T_{-1} := \left(\omega_0 - \Delta\right)/4 = \omega_{-1}/4.
\label{Tlow}
\ea
If higher sidebands ($|m| \geq 2$) are taken into account,
the results remain valid,  except for a  marginal increase  of the relative error bound $\xi$ at
low temperatures (see Apps. \ref{B}-\ref{E}).

By contrast, in the absence of any control ($\mu = 0$), $\varrho_n$ is still given by Eq. \eqref{Treg}, with the average phonon occupation number 
\ba
N_{\rm avg} = N_0 = \frac{1}{e^{\omega_{0}/T} - 1},
\label{eqN0}
\ea
which vanishes in the limit $T \ll \omega_0$. The corresponding QFI is given by
\ba
\mathcal{H}(\mu = 0) = \mathcal{H}_{0}(\mu = 0) = \frac{e^{\frac{\omega_0}{T}} \omega_0^2}{\left(-1+e^{\frac{\omega_0}{T}}\right)^2 T^4},
\label{qfium}
\ea
which has a single peak at $T = T_0$, where $T_0$ satisfies the equation $T_0 = \omega_0 \coth\left(\omega_0/2T_0 \right)/4$ (Cf. inset of Fig. \ref{figqfi}).

One can infer from the above results that a weak sinusoidal modulation (small $\mu$) with appropriately large $\Delta$ 
results in a double-peaked QFI,
 attaining maxima at 
$T \approx T_{-1}$ and $T \approx T_0$, owing to contributions from $\mathcal{H}_{-1}$ and $\mathcal{H}_{0}$ respectively (Fig. \ref{figqfi}).  This double peaked QFI
signifies the applicability of DCQT for estimating a 
much broader range of temperatures, viz., in the vicinity of $T = T_{-1}$, which can be tuned according to our temperature of interest by tuning $\omega_{-1}$, and 
$T = T_0$. In comparison, an uncontrolled quantum thermometer can accurately measure temperatures only in the vicinity of $T = T_0$ for probes characterized by a single energy gap, or 
at multiple fixed (untunable) temperatures, for probes characterized by highly degenerate multiple energy-levels with arbitrary spacings  \cite{campbell18precision}.

For a sub-Ohmic bath spectrum with a low-frequency edge $\omega_{\rm min}$, the minimum error bound $\xi(T = T_{-1})$ per measurement
($\mathcal{M} = 1$) remains finite and approximately constant at (see App. \ref{appD})
\ba
\lim_{\Delta \uparrow \omega_0} \xi(T = T_{-1} \gtrsim \omega_{\rm min}/4) \to  \frac{\exp\left(2\right)}{2\mu},
\label{errT0}
\ea
where $\Delta \uparrow \omega_0$ signifies $\Delta$ approaching $\omega_0$ from below.
Condition \eqref{errT0} yields the maximum possible advantage offered by our control scheme close to the absolute zero (i.e., for $T \approx T_{-1} \ll \omega_0$), and is valid in the regime 
\ba
G_0 \ll \omega_{-1} \sim T;~~\omega_{\rm min} < \omega_{-1} \sim T.
\label{G0llT}
\ea

An error bound $\xi$ which does
not diverge at low temperatures $T \ll \omega_0$ (as in standard quantum thermometry), but instead stays constant as the temperature decreases in that range, is 
deemed to be highly advantageous, since it allows us to 
measure such low temperatures. We have thus obtained a remarkable result: the  advantage offered by our control scheme, expressed by the bound Eq. \eqref{errT0} close to the absolute zero (i.e., for $T \ll \omega_0$), is
 maximal for a sub-Ohmic, nearly-flat 
 bath spectrum (NFBS)
 with small, but non-zero, lower edge, such that the bath interacts weakly
with the DCQT at all non-zero frequencies.
Regime \eqref{G0llT} ensures that the dynamics is Markovian (see Methods). 

Conditions \eqref{errT0}, \eqref{G0llT} allow us to achieve thermometry with a high-precision bound for very low temperatures; viz., only a finite number of measurements
$\mathcal{M} \gg e^4/(4\mu^2)$
ensures a relative error bound which is constant, i.e., temperature-independent, and has the finite value $\exp\left(2\right)/2\mu\sqrt{\mathcal{M}}$.
For  given small values of $G_0$ and $\omega_{\rm min}$, we then arrive at
a minimum (limiting) temperature bound that is 
measurable with the error bound \eqref{errT0} per measurement:
\ba
T &=& {\rm max}\left[T_{\rm lim1}, T_{\rm lim2} \right];\non\\
T_{\rm lim1} &\gg& G_0;~~~ T_{\rm lim2}  > \omega_{\rm min}, 
\label{Tlim}
\ea
  In particular, for a sub-Ohmic NFBS (Eq. \eqref{gnfbs}) coupled very weakly with the thermometer at all frequencies such
that $G_0 \ll \omega_{\rm min} \to 0$, one has $T_{\rm lim} \to 0$, thus enabling us (at least under ideal circumstances, i.e., in absence of any other source of error)
to accurately measure temperatures 
very close to the absolute zero. 

The only penalty for a small $G_0 \ll \omega_{\rm min}$ is long thermalization time.
Yet, even if $G_0$ corresponds to divergent thermalization times (e.g., for $G_0, \omega_{\rm min} \to 0$),
one can  use optimal control to reach the steady state at the minimal (quantum speed limit) time \cite{sugny10}: one can initially 
apply a strong brief pulse to take the thermometer to an optimal state, from which it takes the least time to thermalize, as detailed in \cite{mukherjee13}. 
Following this initial pulse, one can periodically modulate the system, as presented here. 
This behavior  is in sharp contrast to that of an unmodulated 
thermometer, where $\xi(T)$ diverges 
for $T \to 0$, thus precluding any possibility of precise temperature estimation at very low temperatures \cite{potts19fundamental}. 

This novel effect of keeping $\xi$ finite even in close proximity to the absolute zero is a direct consequence of our control scheme, and is unachievable without modulation for any non-zero $\omega_0$. 
Extension of the analysis to multipeak QFI is given in the App. \ref{appF}  (also see Fig. (\ref{fig3p})).

As discussed above, optimal thermometry demands choosing a $\Delta$ such that the QFI has a maximum at the temperature regime of interest, which is always possible for small enough $\omega_{\rm min}$ 
(cf. Eq. \eqref{Tlim}). 
However, even for a bath spectrum with small but finite $\omega_{\rm min}$, one can perform low-temperature thermometry with a low relative error bound in the
regime $G_0\ll T < \omega_{\rm min}$ via sub-optimal DCQT 
whose response is peaked at a temperature close to, but larger than $\omega_{\rm min}$.
In this case the 
QFI, albeit not maximal,  would be still significantly higher than that of the same thermometer in the absence of dynamical control. For example, 
in order to measure temperatures of the order $T \sim 10^{-3}\omega_0$, 
ideally one would need $\omega_m \sim 10^{-3} \omega_0$. However, even a suboptimal modulation with
 $\omega_m \sim 10^{-2} \omega_0$, which leads to QFI with a maximum at $T \sim \omega_m \sim 10^{-2} \omega_0$, results in $\mathcal{H} \approx 14.3$ at $T = 10^{-3}\omega_0$ (see inset of Fig. (\ref{fig3p})).
In contrast, $\mathcal{H}$ is vanishingly small (i.e., $\mathcal{H} \ll 1/T^2$ such that $\xi \gg 1/\sqrt{\mathcal{M}}$; Cf. Eq. \eqref{errb}) for the same temperature 
regime in the absence of control, thus exhibiting the advantage of our DCQT even for suboptimal thermometry at $T < \omega_{\rm min}$.

\begin{figure}[h]
\begin{center}
\includegraphics[width= \columnwidth, angle = 0]{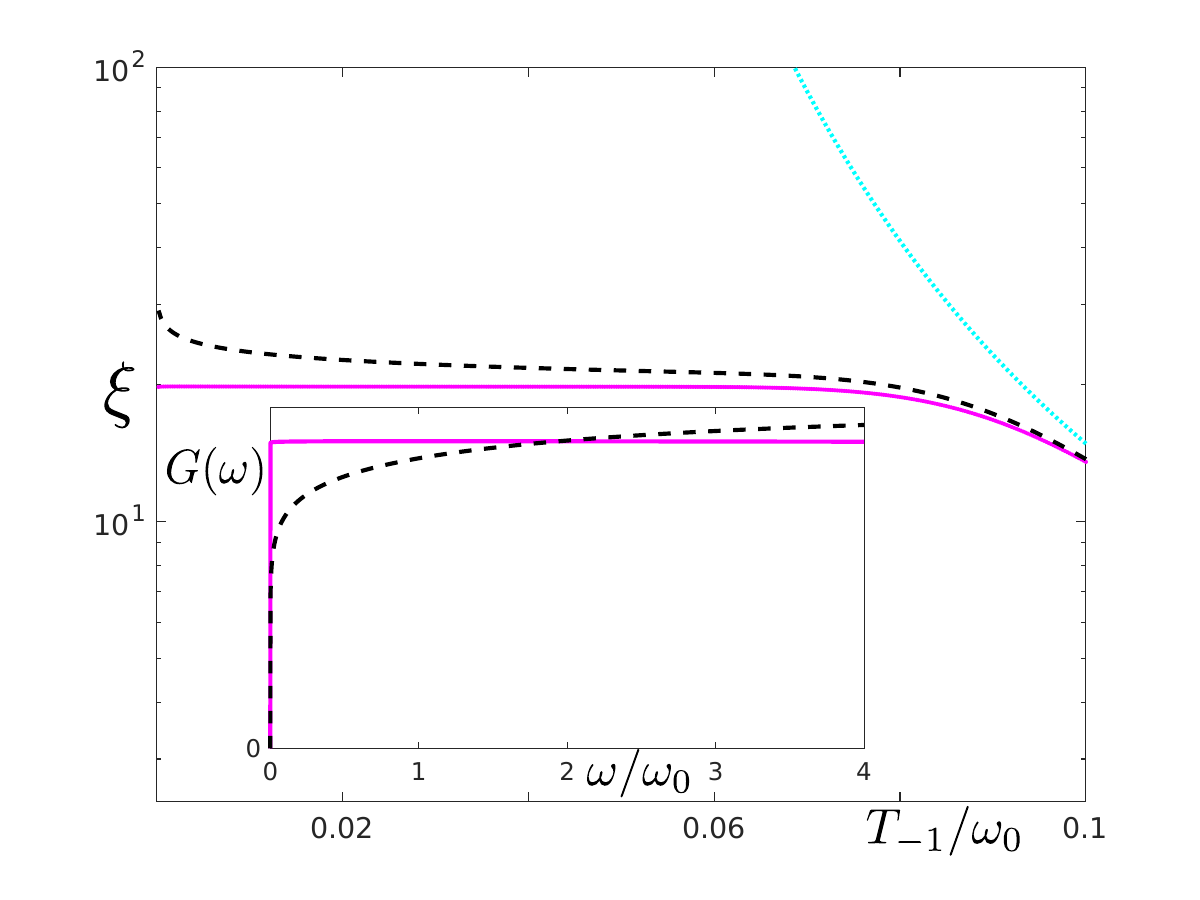}
\end{center}
\caption {{\bf Temperature-independent relative error bound with dynamical control:} Relative error bound $\xi$ for the estimation of bath temperature at $T = T_{-1} = (\omega_0 - \Delta)/4$,  
by a harmonic oscillator dynamically controlled quantum thermometer under sinusoidal modulation (Cf. Eq. \eqref{sinmod}), for the following bath spectra: nearly flat bath
spectrum (magenta solid curve),  
sub-Ohmic bath spectrum with $s = 0.1, \omega_c = 100$ (black dashed curve) and the same spectra in the absence of control (turquoise dotted curve).
Dynamical control reduces the relative error bound significantly,
with $\xi$ remaining constant over a broad range of low temperatures $T \ll \omega_0$.  The sidebands $m = 0, \pm 1, \pm 2, \pm 3$ are included in the calculation. Here  
the modulation amplitude $\mu = 0.2$, and $\omega_0 = 1$.
Inset: Corresponding spectral response functions scaled by $\gamma$.}
\label{figerrgw}
\end{figure}
\begin{figure}[t]
\begin{center}
\includegraphics[width=\columnwidth, angle = 0]{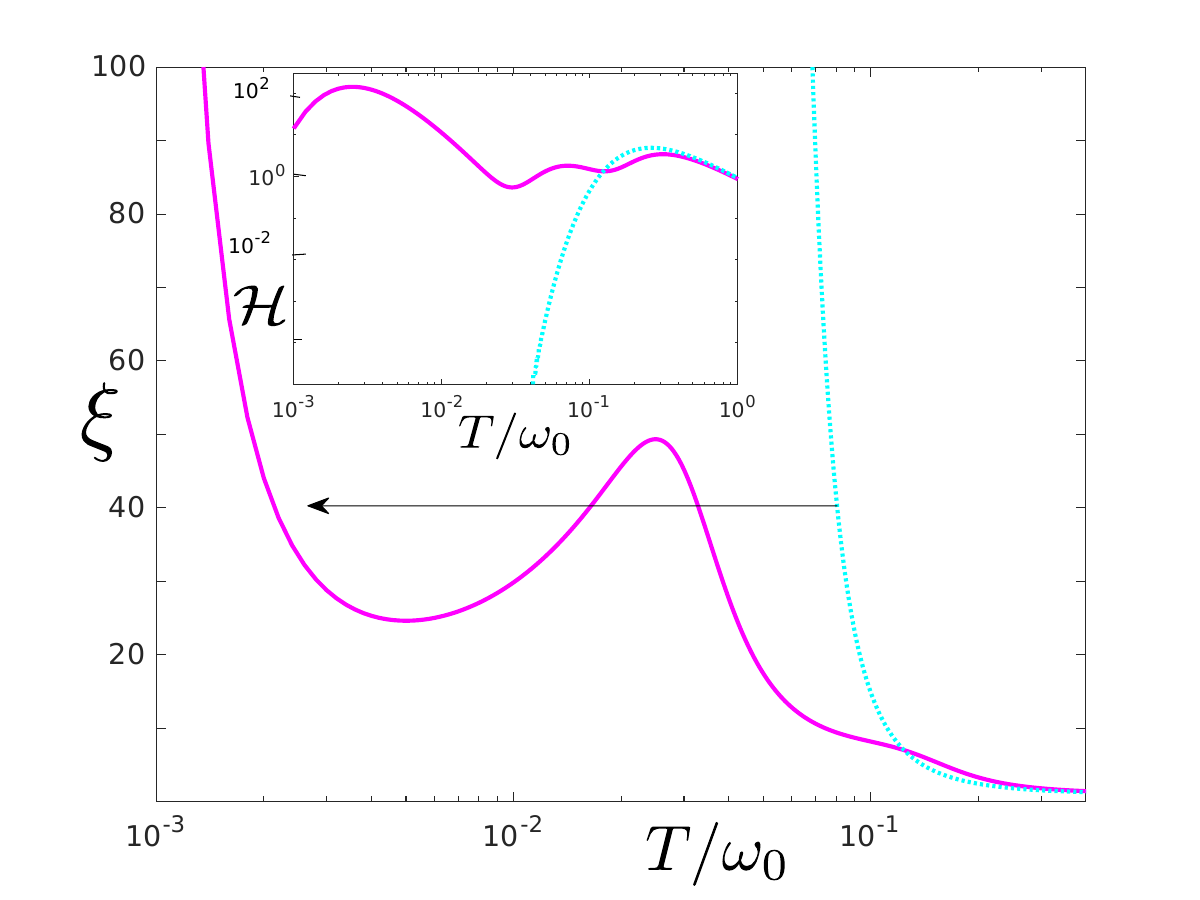}
\end{center}
\caption {{\bf Dynamical control with multiple harmonics:} Relative error bound $\xi$ and  Quantum Fisher Information $\mathcal{H}$ (in inset)   as a function of temperature $T$ (in units of $\omega_0$)
in the absence of any modulation (turquoise dotted curve) and under multi-harmonic  modulation 
$\omega(t) = \omega_0 + \Delta\left(\mu_{l}\sin\left(lt \Delta\right) + \mu_{l^{\prime} \neq l}\sin\left(l^{\prime}t \Delta\right)\right)$ 
 (magenta solid curve), for harmonic oscillator dynamically controlled quantum thermometer probing a bath with nearly-flat bath spectrum. Our control scheme reduces the lowest temperatures measurable
 with finite $\xi$ by almost two orders of magnitude (arrow).
 Here $l = 80,~ l^{\prime} = 99,~ \mu_{80} = 0.394,~\mu_{99} = 0.115,~~\Delta = 0.01$, $\omega_0 = 1$, $\omega_{\rm min} \ll 0.01$ and $\mathcal{M} = 1$.}
\label{fig3p}
\end{figure}

\noindent {\bf Thermometry of arbitrary bath spectra:} The advantage offered by DCQT is not restricted to the specific bath spectra considered above. 
In fact, the key result of significant improvement
in low-temperature thermometry by dynamical
control holds for arbitrary bath spectra satisfying the Kubo-Martin-Schwinger  condition (see Eq. \eqref{kmsgen} of Methods), as long as $G(\omega > 0)$ is independent of temperature.
Choosing a control scheme satisfying (see Eq. \eqref{Tlow})
\ba
\omega_{-1} = \kappa T
\label{eqkap}
\ea
for a positive constant $\kappa$ of the order of unity, 
leads to a QFI which diverges quadratically with temperature:
\ba
\mathcal{H} \approx \frac{\eta e^{\kappa} \kappa^2}{\left(-1+\eta e^{\kappa}\right)^2 T^2}.
\label{qfit2}
\ea
The above 
result Eq. \eqref{qfit2},  which is valid for any bath spectra,  arises due to $\omega_{-1}$ being vanishingly small (see Eq. \eqref{eqkap}). The  lowest temperature for which Eq. \eqref{qfit2} is valid is given by the condition
\ba
\omega_{-1} \sim T \gg T_{\rm lim} = G(\omega_{-1}).
\label{condgen}
\ea
As in the case of sub-Ohmic NFBS, condition \eqref{condgen} ensures that the thermalization time is long enough for the secular approximation to remain valid. 
This in turn results in a relative error bound 
\ba
\xi &\approx& \sqrt{\frac{\left(-1+\eta e^{\kappa}\right)^2}{\eta \kappa^2 e^{\kappa}}}\frac{1}{\sqrt{\mathcal{M}}}, \non\\
\eta &=& \frac{P_0G(\omega_0) + P_1\left(G(\omega_{1}) + G(\omega_{-1}) \right)}{P_1G(\omega_{-1})}.
\label{xinoT}
\ea
which is not explicitly dependent on the temperature, even for low temperatures $T \ll \omega_0$  (see Fig. (\ref{figerrgw})). Although
the relative error bound $\xi$ is not explicitly dependent on  temperature,  the QFI and consequently $\xi$ still
depend on the details of the bath spectral function $G(\omega)$ at $\omega = \omega_{-1}$, which in 
turn is determined by the temperature through Eq. \eqref{eqkap}.
The maximum QFI  in this case
 is obtained by modulations satisfying the optimal condition
\ba
P_1 G(\omega_{-1}) \gg  P_0 G(\omega_{0}),  P_1 G(\omega_{1}).
\label{condopt}
\ea
The above condition ensures that the sidebands $m \neq -1$ are insignificant, so as to yield a small relative error bound 
at $T \sim \omega_{-1}$ (see App. \ref{appC}).

Condition \eqref{condopt} shows that  one needs to tailor the control scheme to the bath spectrum at hand. For example, in the case of a bath spectrum characterized by 
$G(\omega_0) \gg G(\omega_{-1})$,    one needs to design 
a modulation with
$P_{-1} \gg P_0$, in order for our control scheme to be beneficial for estimating temperatures $T \sim T_{-1} \ll \omega_0$. This can  be realized using a periodic $\pi$-pulse modulation, in which 
case $P_{\pm 1} \approx 4/\pi^2$ and $P_{m \neq \pm 1} \approx 0$ \cite{klimovsky13}.

\section*{Discussion}
We have shown that by subjecting a generic quantum thermometer to an appropriate  dynamical control, we may increase its maximum accuracy bound
of measuring
a chosen range  of temperatures. The class of quantum thermometers considered here relies on the measurement of the energy-level populations $\varrho_n$ at
the steady-state, as well as knowledge of the resonant frequency $\omega_0$ of the thermometer, the modulation parameters $\mu$ and $\Delta$ (cf. \eqref{sinmod}) and (at least crudely) 
the bath spectral density $G(\omega)$ (see App. \ref{appA}). It falls in the category of secondary thermometers.
The advantage of 
the proposed DCQT becomes especially apparent at low temperatures, where,  for  generic bath spectra $G(\omega)$, its accuracy bound of measuring the bath temperature, quantified by  the  QFI, is 
dramatically higher than that of its unmodulated counterpart. Namely, 
 dynamical control  allows us to perform low-temperature thermometry with temperature-independent  relative error bound, at temperatures above the bound set by Eq. \eqref{Tlim}. Our 
proposed control scheme can be tailored according to the bath spectra at hand, in order to maximize the QFI, and determine the number of its peaks and sensitivity ranges, 
thus making it highly versatile. 

 In addition to our diagonal Hamiltonian $\hat{H}(t) = \sum_{n=0}^{\mathcal{N}} \omega_n(t) \ket{n}\bra{n}$, off-diagonal interaction terms of the 
 form $\sum_{n,m = 0}^{\mathcal{N}} \left(h_{n,m} \ket{n}\bra{m} + h.c. \right)$ may incur  non-adiabatic effects, which may cause spurious transitions between the energy levels, even in the absence of any external bath. 
 In order to avoid such excitations one should require $\omega(t) \gg h_{n,m}$ for all $t, n, m$, thereby ensuring adiabaticity.

The dynamical control studied here has already been realized experimentally, in the context of probing the system-environment coupling spectrum \cite{almog11direct}. 
One can utilize similar dynamical control to realize DCQT in different multilevel systems (see App. \ref{appA}): i) a harmonic oscillator DCQT can be implemented by the mode 
of a  cavity of length $l(t)$, whose frequency $\omega(t) \propto 1/l(t)$
 is modulated with the inverse of the cavity length \cite{macri18nonperturbative}.
ii) A two-level DCQT can be implemented by a qubit whose level spacing is modulated by a time-dependent magnetic field, for example in NMR \cite{prigl96a} and
NV-center \cite{casola18probing} experiments. 
Experimentally, one can probe the thermal steady-state $\hat{\rho}(t\to \infty)$ of the DCQT at large times, 
after decoupling it from the bath, followed by
ceasing the dynamical control (i.e., first setting $\gamma = 0$, and then $\mu = 0$). In a harmonic-oscillator thermometer one can then
estimate the temperature $T$ through measurement of the mean number of quanta of the thermometer $N_{\rm avg}$, for example through
motional sideband spectroscopy, which relies on the asymmetry between phonon absorption (proportional to $N_{\rm avg}$) and phonon emission 
(proportional to $N_{\rm avg} + 1$) \cite{safavi-naeini12observation}.  In a two-level NV-center thermometer \cite{hopper18spin, tran18anti}, spin readout using photoluminescence may enable us to
measure the N-V temperature. 

The thermal energy-level population measurements $\varrho_n$ (see Eq. \eqref{rhoneff}), which constitute the optimal 
positive-operator valued measure for estimating temperature \cite{brunelli11}, are characterized 
solely  by the average number of quanta  $N_{\rm avg}$. Consequently, measurement of $N_{\rm avg}$ gives us the complete knowledge of
$\hat{\rho}(t\to \infty)$, which in turn enables
us to  approach the minimum error bound $\xi$ of bath-temperature estimation. However, while DCQT can dramatically reduce the error bound of temperature 
estimation (see Eqs. \eqref{condgen} and \eqref{xinoT}), additional  
experimental errors may arise from the finite accuracy of measuring $\Delta$, or the energy-level populations. These errors, which depend on the experimental apparatus,  may prevent us 
from reaching the fundamental theoretical bound of Eq. \eqref{xinoT} (see App. \ref{appB}).

DCQT can be highly advantageous 
for thermometry of diverse baths realized by many-body quantum systems in condensed matter and ultracold atomic gases, with spectra and modulations satisfying the optimal condition Eq. \eqref{condopt}.

The temperatures measurable with low relative error bound using our control scheme are presented in Table \ref{tabpar}.
\begin{table}
\label{tabdef}
\begin{center}
\renewcommand{\arraystretch}{2}
\begin{tabular}{ | m{6em} | m{6.5cm}|} 
\hline
Parameters & \hspace{2.4cm}Values \\
\hline
 $\omega_{-1}$ & $\sim 10^{x}$ Hz\\ 
\hline
$T_{-1}$ &  $\approx 1.9 \times 10^{x-12}$ K  \\  
\hline
$\small{\mathcal{H}(T = T_{-1})}$ & $\approx \frac{P_1 e^{\frac{\hbar \omega_{-1}}{k_B T}} \hbar^2 \omega_{-1}^2}{\left(-P_1 + e^{\frac{\hbar \omega_{-1}}{k_B T}}\right)^2 k_B^2 T^2}\frac{1}{T^2} {\rm K}^{-2} \approx \frac{4 \mu^2}{e^4}\frac{1}{T^2} {\rm K}^{-2}$\\  [1ex]
\hline
$\xi(T = T_{-1}) $ & $\approx \exp\left(2\right)/\left(2\mu\sqrt{\mathcal{M}} \right)$ \\ 
\hline
\end{tabular}
\renewcommand{\arraystretch}{1}
\end{center}
\caption{{\bf Illustrative values:} Values of $T_{-1} = \hbar \omega_{-1}/4k_B$, quantum Fisher information $\mathcal{H}(T = T_{-1})$ and relative error bound $\xi(T = T_{-1})$ for 
$\omega_{-1} = \omega_0 - \Delta \sim 10^{x}$ Hz, for a harmonic oscillator dynamically controlled quantum thermometer under sinusoidal modulation probing a nearly-flat bath spectrum. Here 
  $x$ is a real number.}
\label{tabpar}
\end{table}
In a microwave cavity subjected to sinusoidal
modulation with $\omega_0 - \Delta \sim 10^x$ Hz, DCQT 
results in QFI attaining a peak at temperatures of the order of $10^{x-12}$ K, $x$ being real. We thereby open new avenues for the study of cavity quantum electrodynamics and quantum
information processing \cite{hauke16} at extremely 
low temperatures. The caveat noted above is that although the dynamical control can be expected to significantly reduce the error in low-temperature thermometry, however, experimental errors may preclude the attainment of theoretical bound of temperature resolution.

DCQT can be expected to be highly beneficial in baths exhibiting the widely studied $1/f^{\alpha}$ ($\alpha \geq 0$) noise spectra \cite{milotti02onebyf} 
 in vacuum tubes \cite{johnson25the} or in thick film resistors \cite{pellegrini83noise}. This kind of spectra ensures that for a sinusoidal modulation with small $\omega_{-1}$,
the most dominant contribution arises from the $m = -1$ sideband, thus enabling us to probe low frequencies with high precision bound.

 An intriguing application of the proposed DCQT concerns experimentally studying the third law of thermodynamics
for low-temperature many-body quantum systems, in the sense of understanding the scaling of the cooling rate with temperature \cite{masanes17a, freitas17fundamental, cleuren12}. 
In view of predictions that the cooling rate does not vanish 
as the absolute zero is approached for baths with anomalous dispersion,
such as magnon (ferromagnetic) spin chains \cite{kolar12}, high-precision low-temperature thermometry is imperative.

In many-body quantum 
systems \cite{rigol08, eisert15, calabrese11, alvarez15, kaufman16}, interactions 
may keep a system non-equilibrated for a long time after a quench \cite{breuer02}, during which time its different collective modes $k$
may have their own 
temperatures $T_k$.
High-precision multi-mode thermometry over a wide range of temperatures using our DCQT can verify whether different modes have different temperatures, and thus avoid thermalization.

The versatility of our DCQT can be well-suited for
simultaneous multi-mode probing of a bath with high accuracy bound, which can be especially useful for nanometer-scale thermometry in biological systems \cite{kucsko13, baffou14a}. 

The recently discussed need for thermometers with vanishing energy gaps for measuring low-temperatures in many-body quantum systems \cite{potts19fundamental} and in 
strongly coupled quantum systems \cite{hovhannisyan18measuring}, 
suggests the importance of control schemes capable of tuning the gap of a quantum thermometer to our advantage. 
Application of our control scheme to thermometers modelled by many-body quantum
systems, or to thermometers coupled strongly to the bath, in order to achieve high-precision low-temperature thermometry, is an interesting question which we 
aim to address in the future.

The control scheme presented above may have diverse applications in quantum metrology beyond thermometry. For example,
periodic modulation of energy levels in many-body quantum critical 
systems may be applicable to precise probing of inter-particle coupling strengths in these systems \cite{zanardi08quantum}.

\section*{Methods}

\subsection*{Thermalization under periodic modulation}

We consider a multi-mode DCQT system with its state $\rho(t)$ given as a direct product of single-mode states $\rho_k$: $\rho(t) = \otimes_k \rho_k(t)$,
interacting with a multi-mode bath, again described by the direct product state $\rho_{B} = \otimes_k \rho_{B_k}$. Here $\rho_{B_k}$ denotes the state of the $k$-th mode of the bath. 
For each $k$, $\rho_k(t)$  evolves under the action of a periodic Hamiltonian, satisfying Eq. \eqref{hamilN}.
The system is coupled to the bath mode through the interaction Hamiltonian
\ba
\hat{H}_{Ik} = \hat{S}_k\otimes \hat{B}_k,
\label{hintk}
\ea
 where each of the independent $k$-th mode baths  
has a spectrum wide  enough to give rise to Markovian dynamics; for example, a finite-Q cavity mode, or a finite-lifetime phonon mode. In case the above assumption is violated,
leading to non-Markovian dynamics \cite{breuer02, zwick14},  the DCQT may be useful for revealing the absence of thermalization. Here we allow for
baths with mode-dependent temperatures $T_k$, which we aim to measure accurately using our DCQT.

\par 

We sketch below  the derivation of the master equation describing the thermalization of a system under periodic control. We refer to
\cite{breuer02, klimovsky13, klimovsky15, alicki14} 
for details of the derivation, and \cite{almog11direct} for experimental studies of open quantum system in presence of dynamical control. 
(For reviews on periodically driven  open quantum systems, see \cite{klimovsky15, kosloff13}.)

\par

The time evolution operator for the periodic Hamiltonian Eq. \eqref{hamilN} is given by 
\ba
\hat{U}_k(t,0) = \hat{\mathcal{T}}\exp\left(-i\int^{t}_0 \hat{H}_k(t^{\prime}) dt^{\prime} \right),
\label{uper}
\ea
$\hat{\mathcal{T}}$ being the time ordering operator.
According to the Floquet theorem, one can decompose the time evolution operator as $\hat{U}_k(t,0) = \hat{P}_{k}(t)e^{\hat{R}_k t}$, where $\hat{P}_{k}(t + \tau) = \hat{P}_{k}(t)$ and $\hat{R}_k$ is
a constant operator. Taking into account $\hat{U}_k(0,0)= \mathds{1}$, and the periodicity of $\hat{P}_{k}(t)$, one obtains $\hat{P}_k(0)=\mathds{1}$ and hence
$\hat{U}_k(\tau,0)=\hat{P}_k(\tau)e^{\hat{R}_k\tau}=\hat{P}(0)e^{\hat{R}_k\tau}=e^{\hat{R}_k\tau}$. 
One can now identify the   operator $\hat{R}_k$ with an effective Hamiltonian $\hat{H}_{k,\rm eff}$ averaged over a period of $\hat{H}_k(t)$, as   
\ba
\hat{U}_k(\tau,0) = e^{\hat{R}_k \tau} =: e^{-i \hat{H}_{k,\rm eff}\tau},
\ea
with eigenenergies $\Omega_{k,r}$ and eigenstates $\ket{r_k}$, through the relation 
 \ba
 \hat{H}_{k,\rm eff} = \sum_{r} \Omega_{k,r_k} \ket{r_k}\bra{r_k}.
 \label{hper}
 \ea

The Fourier components $\hat{S}_{\omega_k,m}$ of the system operator $ \hat{S}_k(t)$  (Cf. Eq. \eqref{hintk}) in the interaction picture with respect to the evolution \eqref{uper}, are given as
\ba
 \hat{S}_k(t) &=& \hat{U}_k^{\dagger}(t,0) \hat{S}_k \hat{U}_k(t,0) \non\\
 &=& \sum_{\{\omega_k\}}\sum_{m \in \mathbb{Z}} \hat{S}_{\omega_k,m} e^{i\left(\omega_k + m\Delta \right)t},
\ea
where $\Delta = 2\pi/\tau$, $m$ are integers, and $\{\omega_k\}$ is defined as the set of all transition frequencies $\Omega_{k,r^{\prime}} - \Omega_{k,r}$ between the levels of $\hat{H}_{k,\rm eff}$ and
the operators $\hat{S}_{\omega_k,m}$ are the $m$th-harmonic transition operators associated with these levels \cite{klimovsky15}.

\par

Next we focus only on the dynamics of the $k$-th mode of the thermometer and the bath. Under the standard Born-Markov approximation in the weak thermometer-bath coupling limit,
we arrive at the master equation \cite{breuer02}
\ba
&&\frac{d}{dt}\hat{\rho}_k(t) \\
&=& -\int^{\infty}_0 ds {\rm Tr}_{B_k}\Big[H_{Ik}(t),\left[\hat{H}_{Ik}(t-s),\hat{\rho}_k(t)\otimes\hat{\rho}_{B_k}\right]\Big]\non,
\ea
where ${\rm Tr}_{B_k}$ denotes trace over the bath mode $k$, and $\hat{H}_{Ik}(t)$ is the $k$-th mode interaction Hamiltonian in the interaction picture.

We now assume that the thermalization time of the thermometer induced by the bath is much longer than 
$\tau$ or $\omega_{k,m}^{-1} \equiv \left(\omega_k + m\Delta \right)^{-1}$. Consistently, 
 we adopt the secular approximation (SA) to average out 
the rapidly oscillating terms of the form $\exp\left[\pm i\left(\omega_{k,m} - \omega_{k^{\prime},m^{\prime}}\right)t\right]$ at times $t \gg \tau, \omega_{k,m}^{-1}$, 
such that only the 
terms with $k = k^{\prime},~ m = m^{\prime}$ survive \cite{mukherjee16speed}. We then finally arrive
at the master equation \cite{klimovsky13, klimovsky15, alicki14,  breuer02}, which is a weighted 
sum of Lindbladian superoperators, each valid for system-bath coupling centered at $\omega_{k,m} := \omega_k + m\Delta$:
\ba
\dot{\hat{\rho}}_k(t) =  \mathcal{L}_k\left[\hat{\rho}_k (t)\right] = \sum_{\{\omega_k \}}\sum_m\mathcal{L}_{\omega_k,m} \left[\hat{\rho}_k (t)\right].
\ea
Here
\ba
\mathcal{L}_{\omega_k,m} \left[\hat{\rho}_k(t)\right] &=& G(\omega_{k} + m\Delta)\mathcal{D}_{\omega_k,m}[\hat{\rho}_k(t)] \non\\ 
&+&  G(-\omega_{k} - m\Delta) \mathcal{D}_{\omega_k,m}^{\dagger}[\hat{\rho}_k(t)],\non\\
\mathcal{D}_{\omega_k,m}[\hat{\rho}_k(t)] &=& \Big(\hat{S}_{\omega_k,m} \hat{\rho}_k(t) \hat{S}^{\dagger}_{\omega_k,m} \non\\ 
&-& \frac{1}{2}\hat{S}^{\dagger}_{\omega_k,m}\hat{S}_{\omega_k,m}\hat{\rho}_k(t) \non\\
&-& \frac{1}{2}\hat{\rho}_k(t) \hat{S}^{\dagger}_{\omega_k,m}\hat{S}_{\omega_k,m} \Big),
\label{lind}
\ea
and
\ba
&& G(\pm \left(\omega_k + m\Delta\right)) \non\\
&=& \int_{-\infty}^{\infty} e^{\pm i\left(\omega_k + m\Delta\right)t} \langle \hat{B}_k(t)\hat{B}_k(0)\rangle dt\non,
\ea
where $\hat{B}_k(t) = e^{i \hat{H}_{B} t}\hat{B}_ke^{-i \hat{H}_{B}t}$, 
$\hat{H}_{B}$ being the bath Hamiltonian,
represents the bath spectral response or auto-correlation function sampled at the frequency harmonics $\pm \left(\omega_k + m\Delta\right)$. The Kubo-Martin-Schwinger condition must be
imposed, i.e.,
\ba
\frac{G(\omega_k + m\Delta)}{G(-\omega_k - m\Delta)} = \exp\left[\left(\omega_k + m\Delta\right)/T_k\right].
\label{kmsgen}
\ea
 Eqs. \eqref{lind} - \eqref{kmsgen} imply that, within the approximations assumed above, $\rho_k$ equilibrates to a thermal (Gibbs) state at temperature 
$T_k$. The 
level populations are determined by the weighted bath response at the resonance frequencies $\omega_k$ shifted by Floquet harmonics of modulation, $\omega_{k,m} = \omega_{k} + m\Delta$.

\section*{Data availability}
All relevant data are available to any reader upon reasonable request.

\section*{Code availability}
All relevant codes are available to any reader upon reasonable request.

\section*{Acknowledgements}
The authors acknowledge Wolfgang Niedenzu for helpful discussions. The support of DFG (FOR 7024), EU-FET Open (PATHOS) project, ISF, VATAT, NSFC (11474193), CONICET, CNEA, Shuguang (14SG35), STCSM (18010500400 and 18ZR1415500), 
the Program for Eastern Scholar and the Ram\'{o}ny Cajal program (RYC-2017-22482) of the Spanish MINECO, Initiation Grant (2018513) and SRG/2019000289
are acknowledged.  

\section*{Author Contributions}

V.M. and G.K. conceived the idea. V.M. and A.Z. performed the analytical calculations and numerical simulations. V.M., A.Z., A.G., X.C. and G.K. 
were involved  in the interpretation of the results and in the discussions during the writing of the manuscript. V.M. and G.K. wrote the manuscript.

\appendix

\section{$\mathcal{N}$ level dynamically controlled thermometer}
\label{appA}

 A single bath probed by a single-mode $\mathcal{N}$ level system, subject to a (generic) diagonal modulation is described by the Hamiltonian
\ba
\hat{H}(t) = \sum_{n=0}^{\mathcal{N}} \omega_n(t) \ket{n}\bra{n} = \sum_{n=0}^{\mathcal{N}} n \omega (t) \ket{n}\bra{n},
\ea
with 
\ba
\omega_0  = \frac{1}{\tau}\int^{\tau}_0 \omega(t) dt, ~~~~  \hat{H}_{\rm eff} = \sum_{n=0}^{\mathcal{N}} n\omega_0 \ket{n}\bra{n}.
\ea
The interaction Hamiltonian has the form 
\ba
\hat{H}_{I} = \hat{S}\otimes \hat{B},
\ea
where $\hat{S}$ is a system-operator belonging  to a Lie algebra, and the operator $\hat{B}$ acts on the bath.
Under modulation with period $\tau$ the master equation for the state of the system reduces to the Floquet expansion \cite{alicki12periodically, szczygielski13, kosloff13, klimovsky13, klimovsky115laser, klimovsky15, alicki14,  breuer02, carmichael02, agarwal13, alicki18introduction,  kofman01universal, kofman04unified, gordon07universal, clausen10bath, shahmoon13, zwick14, mukherjee16speed, sillanpaa09autler} 
\ba
\dot{\hat{\rho}}(t) &=& \sum_m \mathcal{L}_m\left[\hat{\rho} (t)\right]\non\\
\mathcal{L}_{m}[\hat{\rho}(t)] &=& \left[G(\omega_m)\mathcal{D}_m[\hat{\rho}(t)] +  G(-\omega_m)\mathcal{D}_m^{\dagger}[\hat{\rho}(t)]\right]\non\\
\mathcal{D}_m[\hat{\rho}] &=& P_m \mathcal{D}[\hat{\rho}] \non\\&=& \Big(\hat{S}_{m}\hat{\rho} \hat{S}_{m}^{\dagger} - \frac{1}{2}\hat{S}_{m}^{\dagger}\hat{S}_{m}\hat{\rho} - \frac{1}{2}\hat{\rho} \hat{S}_m^{\dagger}\hat{S}_{m} \Big)\non\\
\hat{S}_m &=& P_m \hat{S}_{-};~~\hat{S}_m^{\dagger} = P_m \hat{S}_{+}.
\label{megen}
\ea
\begin{figure}[h]
\begin{center}
\includegraphics[width= \columnwidth, angle = 0]{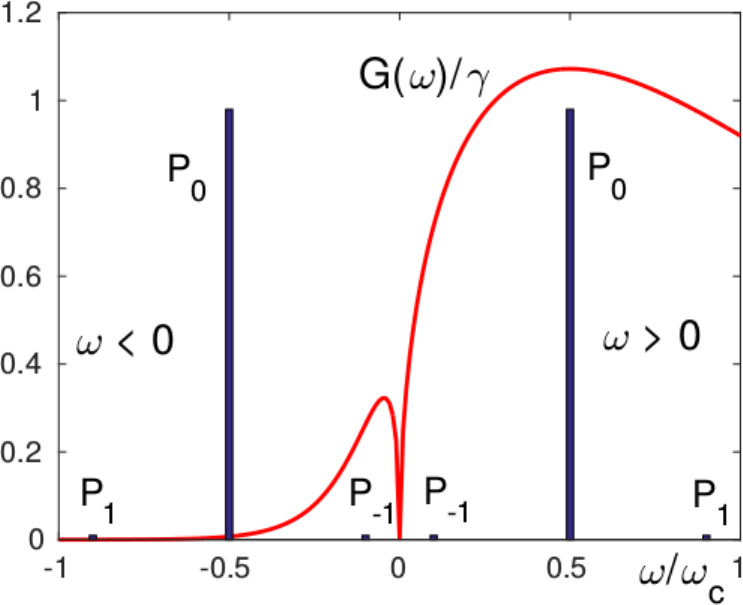}
\end{center}
\caption {{\bf Floquet sidebands:} A sub-Ohmic bath spectral function  $G(\omega > 0) = \gamma \left(\omega^s/\omega_c^{s-1}\right)\exp\left[-\omega/\omega_c\right]$ (scaled by $\gamma$) (red curve); $\gamma$ determines 
the amplitude of $G(\omega)$. According to the Kubo-Martin-Schwinger condition,
$G(-\omega) = G(\omega)\exp\left(-\omega/T \right)$. The weights $P_m$ at frequencies $\omega_m = \omega_0 + m\Delta$ are shown for a weak  sinusoidal modulation (Eq. \eqref{omtsup}) for 
$m = 0, \pm 1$ (blue lines). As seen above, $P_{\pm 1} \ll P_0$. 
The higher sidebands ($|m| > 1$) are associated with negligible weights $P_m$'s, and can be neglected in this regime.
Here $s = 0.5,~ \omega_c = 1,~ \omega_0 = 0.5\omega_c,~
T = 0.1\omega_c, ~\Delta = 0.4\omega_c$ and $\mu = 0.2$. }
\label{figgw}
\end{figure}
Here the Floquet sideband frequencies labeled by integer $m$ are $\omega_m = \omega_0 + m\Delta$, $\Delta = 2\pi/\tau$, $P_m = P_{-m}$ denotes the weight of the 
$m$-th sideband and $\hat{S}_{\pm}$ are the corresponding 
system excitation and deexcitation (ladder) operators. A physical bath has spectral response $G(\omega)$ that is non-zero and finite for $\omega > 0$, 
satisfies $G(\omega \to 0) \to 0$. Here we consider $G(\omega > 0)$ to be independent of temperature, while 
the Kubo-Martin-Schwinger (KMS) condition (Cf. Eq. (25) of main text), determines $G(\omega < 0)$ through the relation\cite{breuer02}: 
\ba
G(-\omega) = G(\omega) e^{-\omega/T}.
\label{kmsN}
\ea
 
For such bath-response spectra one can rewrite Eq. \eqref{megen} as
\ba
\dot{\hat{\rho}}(t) &=& \sum\nolimits_{m} P_m \Big[G(\omega_m)\mathcal{D}[\hat{\rho}(t)] +  G(-\omega_m)\mathcal{D}^{\dagger}[\hat{\rho}(t)]\Big] \non\\
&=& \Big(\sum\nolimits_{m}^+ P_m G(\omega_m) \non\\ &+& \sum\nolimits_{m}^- P_m G(|\omega_m|)e^{-|\omega_m|/T}\Big)\mathcal{D}[\hat{\rho}(t)]\non\\ &+& \Big(\sum\nolimits_{m}^+ P_m G(\omega_m)e^{-\omega_m/T} \non\\ &+& \sum\nolimits_{m}^- P_m G(|\omega_m|)\Big)\mathcal{D}^{\dagger}[\hat{\rho}(t)].
\label{megenexp}
\ea
Here $\sum_m$ denotes the Floquet summation over all integers $m$, while $\sum^{+}_m$ ($\sum^{-}_m)$ denotes summation over integer $m$ such that $\omega_m \geq 0$ ($\omega_m < 0$).

As the thermometer thermalizes with the bath, its state converges to the thermal (Gibbs) state upon neglecting the fast oscillating terms:
\ba
\hat{\rho}(t \to \infty) &=& \sum_{n = 0}^{\mathcal{N}} \varrho_{n} \ket{n}\bra{n};\non\\
\varrho_{n} &=& \frac{e^{-n \omega_{\rm eff}/T}}{\sum_{n=0}^{\mathcal N} e^{-n \omega_{\rm eff}/T}},
\label{apprhoeff}
\ea
where the probability ratios (Boltzmann factors) of levels $n$ and $n+1$ satisfy
\ba
\frac{\varrho_{n+1}}{\varrho_{n}} &=& e^{-\omega_{\rm eff}/T} \non\\&=&
 \frac{\sum_{m} G(-\omega_m) P_{m}}{\sum_{m} G(\omega_m) P_m};~~ m = 0, \pm 1, \pm 2, \ldots
\ea
Taking into account the definition of $G(\omega)$ and the KMS condition, we get the general Floquet form of the Boltzmann factors
\ba
e^{-\omega_{\rm eff}/T} = \frac{\sum^{+}_{m} G(\omega_m) P_m e^{-\omega_m/T} + \sum^{-}_{m} G(|\omega_m|) P_m}{\sum^{+}_{m} G(\omega_m) P_m + \sum^{-}_{m} G(|\omega_m|) P_me^{-|\omega_m|/T}}.
\label{appboltzeff}
\ea

From Eq. \eqref{appboltzeff} for the asymptotic thermalized populations, we may calculate the quantum Fisher Information (QFI), which is defined as \cite{correa15, paris09}
\ba
\mathcal{H} &=& -2\lim_{\epsilon \to 0}\partial^2 F(\hat{\rho}(T,t), \hat{\rho}(T+\epsilon,t))/\partial \epsilon^2 \non\\&=& \sum_{n = 0}^{\mathcal{N}} \frac{|\frac{\partial\varrho_{n}}{\partial T}|^2}{\varrho_{n}},
\label{qfignsn}
\ea
where $F(\hat{\rho}_1, \hat{\rho}_2) = {\rm Tr}\left[\sqrt{\sqrt{\hat{\rho}_1}\hat{\rho}_2\sqrt{\hat{\rho}_1}}\right]$ is the 
fidelity between $\hat{\rho}_1$ and $\hat{\rho}_2$. 
This expression can be shown to assume the generic form
\ba
\mathcal{H}  = \sum_{m}f(\{G_m\}, \{P_m\}, \{\exp\left(-|\omega_m|/T\right)\}, \frac{\{\omega_m\}}{T^2}, \mathcal{N}).
\label{qfigenf}
\ea
Here $\{x_m\}$ denotes the set of all $x_m$, and $f$ is an analytic functional form.

In the regime of weak modulation (see App. \ref{appB} below) and $\Delta < \omega_0$,
all the leading  harmonics contributing significantly to the dynamics are associated with $\omega_{m} > 0$, and one can 
neglect all terms with $\omega_m < 0$.  In Fig. (\ref{figgw}) we show as an example a sub-Ohmic bath spectrum satisfying the KMS condition, and the $P_{m}$'s due to a weak sinusoidal modulation  
(see Eq. \eqref{omtsup} below). Eq. \eqref{appboltzeff} then reduces to
\ba
\frac{\varrho_{n+1}}{\varrho_{n}} = e^{-\omega_{\rm eff}/T} \approx \frac{\sum^{+}_{m} G(\omega_m) P_m e^{-\omega_m/T}}{\sum^{+}_{m} G(\omega_m) P_m}.
\label{appboltzeffpos}
\ea
It is seen that the $m$-th sideband contributes significantly to the Boltzmann factors in the steady state for $\omega_m \lesssim T$, provided $P_m G(\omega_m)$, 
also known as the 
spectral-filter values \cite{kofman01universal, kofman04unified}, is large enough. The explicit form of the QFI is obtained from Eqs. \eqref{qfignsn} and \eqref{appboltzeffpos} as
\ba
\mathcal{H} \approx \frac{A_{\rm gen}}{B_{\rm gen}},
\label{qfignsnAB}
\ea
where,
\begin{widetext}
\[
A_{\rm gen} = \sum_{n=0}^{\mathcal{N}} \left(\left(\frac{\kappa}{\zeta}\right)^{-2+n}
\left(-\kappa\sum_{n=0}^{\mathcal{N}} \frac{n \left(\frac{\kappa}{\zeta}\right)^{-1+n} C}{\zeta}  
+ n \left(\sum_{n=0}^{\mathcal{N}} \left(\frac{\kappa}{\zeta}\right)^n\right)C\right)^2\right),
\]
\end{widetext}
\ba
B_{\rm gen} = T^4 \zeta^2 \Big(\sum_{n=0}^{N} \Big(\frac{\kappa}{\zeta}\Big)^n\Big)^3, \non
\ea
and
\begin{gather}
C = \sum_{m} e^{-\frac{\omega_m}{T}} P_m G_m  \omega_m; \quad 
\kappa = \sum_{m} e^{-\frac{\omega_m}{T}} P_m G_m; \non\\
\zeta = \sum_{m}  P_m G_m; \quad \quad \quad
\omega_m > 0.
\end{gather}
As shown for specific examples in the main text, as well as in Apps. \ref{appB} - \ref{appD}, the QFI in  Eq. \eqref{qfigenf} can be strongly enhanced at specific $T$ by an appropriate choice of the leading 
Floquet harmonics (sidebands) for a given bath response spectrum (see Fig. \ref{figm1}a for the QFI dependence on modulation frequencies and bath temperatures, 
considering the zeroth and the
first three sidebands). Because of the restriction $\sum_m P_m = 1$, in general only a few of the harmonics, which 
correspond to the frequency of the unmodulated system shifted by low multiples of the modulation frequency $\Delta$, contribute to the expressions for $A_{\rm gen}$, $B_{\rm gen}$, and thus to the QFI.

\section{Generic bath spectrum}
\label{appB}

Here we focus on a harmonic oscillator dynamically controlled quantum thermometer (DCQT) under a weak (small-amplitude) sinusoidal modulation
\ba
\hat{H}(t) &=& \omega(t)\hat{a}^{\dagger}\hat{a} = \sum_{n=0}^{\infty} n \omega (t) \ket{n}\bra{n},\non\\
\omega(t) &=& \omega_0 + \mu \Delta \sin \left(t \Delta \right),
\label{omtsup}
\ea
with $0 < \mu \ll 1$.

For $\hat{S} = \left(\hat{a} + \hat{a}^{\dagger}\right)$, the interaction Hamiltonian becomes
\ba
\hat{H}_{I} = \left(\hat{a} + \hat{a}^{\dagger}\right)\otimes \hat{B}.
\ea
In the interaction picture, we have
\ba
\hat{S}(t) &=& \hat{U}^{\dagger}(t,0) \hat{S} \hat{U}(t,0) \non\\&=& \exp\left[{-i\int^{t}_0\omega (t^{\prime}) \hat{a}^{\dagger}\hat{a} dt^{\prime} }\right] \hat{S} \non\\ &&\exp\left[{i\int^{t}_0\omega (t^{\prime}) \hat{a}^{\dagger}\hat{a} dt^{\prime} }\right],
\label{su}
\ea
which one can expand as
\ba
\hat{S}(t) &=& \sum_m \left[\varepsilon_m e^{-i\left(\omega_0 + m\Delta\right)t}\hat{a} + \varepsilon_m^{*} e^{i\left(\omega_0 + m\Delta\right)t}\hat{a}^{\dagger} \right],\non\\ m &=& 0, \pm 1, \pm 2, \pm 3, \hdots .
\label{sufourier}
\ea
From Eqs. \eqref{su} and \eqref{sufourier}, one can show that the harmonic weights are given by \cite{alicki12periodically, klimovsky13, klimovsky15}
\ba
P_m &=& |\varepsilon_m|^2 = \left|\frac{1}{\tau}\int^{\tau}_0 e^{-i\int_0^{t} \left(\omega(t^{\prime}) - \omega_0\right) dt^{\prime}}e^{i t \Delta} dt \right|^2 \non\\ &=& P_{-m}.
\label{pmsup}
\ea
The weights $P_m$ can be evaluated in terms of Bessel functions.  Since we assume that the amplitude $\mu \ll 1$, the lowest order terms 
are given by \cite{klimovsky13, alicki14, klimovsky15}
\ba
P_0 &\approx& 1-\frac{\mu^2}{2} + \frac{3\mu^4}{32} - \frac{5\mu^6}{576} + \mathcal{O}\left(\mu^8\right), \non\\
P_{\pm 1} &\approx& \frac{\mu^2}{4} - \frac{\mu^4}{16} + \frac{5\mu^6}{768} + \mathcal{O}\left(\mu^8\right), \non\\
P_{\pm 2} &\approx& \frac{\mu^4}{64} - \frac{\mu^6}{384}  + \mathcal{O}\left(\mu^8\right), \non\\
P_{\pm 3} &\approx& \frac{\mu^6}{2304}  + \mathcal{O}\left(\mu^8\right).
\ea
These weights decrease rapidly with increasing harmonic index $m$.

Unless otherwise stated, we consider below the limit $\Delta \uparrow \omega_0$, where
we define $A\uparrow B$ as $A$ approaching $B$ from below. 
In this case $\omega_m < 0$ for $m \leq -2$, while $\omega_m > 0$  for $m \geq -1$. The master equation \eqref{megenexp} is then evaluated to be
\ba
\dot{\hat{\rho}}(t) &\approx& C_1\left(\hat{a}\hat{\rho}(t)\hat{a}^{\dagger} - \frac{1}{2}\hat{a}^{\dagger}\hat{a}\hat{\rho}(t) - \frac{1}{2}\hat{\rho}(t)\hat{a}^{\dagger}\hat{a} \right) \non\\&+& C_2\left(\hat{a}^{\dagger}\hat{\rho}(t)\hat{a} - \frac{1}{2}\hat{a}\hat{a}^{\dagger}\rho(t) - \frac{1}{2}\hat{\rho}(t)\hat{a}\hat{a}^{\dagger}\right), 
\label{meho}
\ea
where
\begin{widetext}
\begin{equation}
 C_1 = \Big[G(\omega_0)P_0 + P_1\left(G(\omega_1) + G(\omega_{-1}) \right) + P_2\left(G(\omega_2) + G(|\omega_{-2}|)e^{-|\omega_{-2}|/T}\right) + P_3\left(G(\omega_3) + G(|\omega_{-3}|)e^{-|\omega_{-3}|/T}\right)\Big]
\end{equation}
and 
\be
\begin{aligned}
C_2 = \Big[G(\omega_0)P_0 e^{-\omega_0/T} + P_1\left(G(\omega_1)e^{-\omega_{1}/T} + G(\omega_{-1})e^{-\omega_{-1}/T}\right) + P_2 \left(G(\omega_{2})e^{-\omega_{2}/T} + G(|\omega_{-2}|)\right) \\
+  P_3\left( G(\omega_{3})e^{-\omega_{3}/T} + G(|\omega_{-3}|)\right)\Big].
\end{aligned}
\ee
\end{widetext}
Here we have neglected terms of the order $\mathcal{O}\left(\mu^8\right)$, and involved the KMS condition (see main text) $G(-\omega) = G(\omega)\exp\left(-\omega/T\right)$. Equivalent 
forms of the master equation 
\eqref{meho} can be written for other limits of $\Delta$ obeying Eq. \eqref{megenexp}.
As before, we assume the thermalization time ($\sim G(\omega)^{-1}$) of the DCQT is much longer than $\tau$, or $\omega_m^{-1}$, such that the secular approximation is valid. 
 One can write
down the steady state in the form of Eqs. \eqref{apprhoeff} - \eqref{appboltzeff}. 

One can simplify the analysis by noting that for $\mu \to 0$, the higher sidebands do not contribute significantly to the 
QFI, compared to $m = 0,\pm 1$, as also verified numerically by comparing results for different numbers of sidebands (see Fig. \ref{figm1}b).
The average occupation number $N_{\rm eff}$ of the DCQT in thermal equilibrium with the bath has then contributions
from the three harmonics ($m = 0, \pm 1$):
\begin{widetext}
\be
\begin{aligned}
N_{\rm eff} &=& \frac{1}{e^{\omega_{\rm eff}/T} - 1}, \quad \quad \quad e^{-\omega_{\rm eff}/T} &=& \frac{P_0G(\omega_0)e^{-\omega_0/T} + P_1\left(G(\omega_{-1})e^{-\omega_{-1}/T} + G(\omega_1)e^{-\omega_{1}/T}\right)}{P_0 G(\omega_0) + P_1\left(G(\omega_{1}) + G(\omega_{-1}) \right)},
\label{neffgenho}
\end{aligned}
\ee
\end{widetext}
and the $n$-th level occupation probability is \cite{breuer02}
\ba
\varrho_n = \frac{1}{N_{\rm eff} + 1}\left(\frac{N_{\rm eff}}{N_{\rm eff} + 1} \right)^n.
\ea
Here $\omega_0, \omega_{\pm 1} > 0$, as noted above. 

Under the condition $\Delta \uparrow \omega_{0}$, we obtain a small $\omega_{-1}$. In this regime, 
for $ T \sim \omega_{-1} \ll \omega_0, \omega_1$, we get $\exp\left(-\omega_0/T \right), \exp\left(-\omega_{1}/T \right) \to 0$; consequently, we have
\ba
N_{\rm eff} &\approx& \frac{1}{\eta e^{\omega_{-1}/T} - 1},\non\\
\eta &=& \frac{P_0G(\omega_0) + P_1\left(G(\omega_{1}) + G(\omega_{-1}) \right)}{P_1G(\omega_{-1})} \non\\&\geq& 1,
\label{etapg}
\ea
which leads to (see Eq. \eqref{qfignsn})
\ba
\mathcal{H} \approx \frac{\eta e^{\frac{\omega_{-1}}{T}} \omega_{-1}^2}{\left(-1+\eta e^{\frac{\omega_{-1}}{T}}\right)^2 T^4}.
\label{Hal}
\ea

Further, for measuring temperatures of the order of $\sim~T$, one can choose a control scheme with $\Delta = \omega_0 - \kappa T$, such that
\ba
\omega_{-1} = \kappa T,
\label{kap}
\ea
where $\kappa$ is a constant of the order of unity. In this case one gets
\ba
\mathcal{H} \approx \frac{\eta e^{\kappa} \kappa^2}{\left(-1+\eta e^{\kappa}\right)^2 T^2} \sim 1/T^2,
\ea
which finally leads to a relative-error bound in the estimation of $T$ \cite{correa15, paris09}
\ba
\xi = \frac{1}{T\sqrt{\mathcal{M}\mathcal{H}}} = \sqrt{\frac{\left(-1+\eta e^{\kappa}\right)^2}{\eta \kappa^2 e^{\kappa}}}\frac{1}{\sqrt{\mathcal{M}}},
\label{xinoT}
\ea
where $\mathcal{M}$ denotes the number of measurements.
We emphasize that the above relative error bound $\xi$  (Eq. \eqref{xinoT}) has no explicit dependence on temperature, and is valid for arbitrary bath spectra. The control scheme 
enables us to measure ultra-low temperatures with high precision, as long as
$\eta$ in Eq. \eqref{etapg} is not large. However,  even though $\xi$ has no explicit temperature dependence in Eq. \eqref{xinoT}, $\eta$
is a function of $G(\omega_{\pm 1})$, 
while
$\omega_{\pm 1}$ depends on temperature through Eq. \eqref{kap}. This implicit temperature dependence becomes negligible for a $G(\omega)$ varying 
weakly with $\omega$, for example, in the case of a NFBS
(see App. \ref{appD}). On the other hand, for a bath spectrum strongly varying with $\omega$ over a certain frequency interval, for example: 
\ba
G(\omega) &=& \gamma/\omega^{\alpha}\quad\quad \alpha > 0;\non\\G(-\omega) &=& G(\omega) e^{-\omega/T},
\ea
one would require a sufficiently small $\gamma$, such that $G(\omega_{-1} = \kappa T) = \gamma/(\kappa T)^{\alpha}$ is small enough to ensure that the secular approximation is valid. 

One can estimate $T$ by experimentally measuring the average phonon population $N_{\rm eff}$ through the relation (see Eq. \eqref{etapg})
\ba
T \approx \frac{\omega_{0} - \Delta}{\ln \left[\frac{N_{\rm eff} + 1}{\eta N_{\rm eff}} \right]}.
\ea
We note that $\xi$ in Eq. \eqref{xinoT} denotes the fundamental lower bound of the relative error arising due to reduction in the
rate of change of the state $\hat{\rho}(t \to \infty)$ of the DCQT, under a change in $T$, for $T \ll \omega_0$. However, additional sources of error 
can originate from classical uncertainty in the values of the different parameters involved. For example, 
uncertainties $\pm \delta_{\Delta}$ in the modulation frequency $\Delta$ and $\pm \delta_{N_{\rm eff}}$ in the estimation of the phonon population $N_{\rm eff}$  results in the following 
error $\delta T_{\Delta}$ in the estimation of $T$:
\begin{widetext}
\ba
T \pm \delta T_{\Delta} &\approx& \frac{\omega_{-1} \pm \delta_{\Delta}}{\ln \left[\frac{N_{\rm eff} \pm \delta_{N_{\rm eff}} + 1}{\eta N_{\rm eff} \pm \eta\delta_{N_{\rm eff}}} \right]} 
\approx \frac{\omega_{-1}}{\ln\left[\frac{N_{\rm eff} + 1}{\eta N_{\rm eff}} \right]} \pm \frac{\delta_{N_{\rm eff}}}{N_{\rm eff}\left(N_{\rm eff} + 1 \right)\ln\left[\frac{N_{\rm eff} + 1}{\eta N_{\rm eff}} \right]} \cdot \frac{\omega_{-1}}{\ln\left[\frac{N_{\rm eff} + 1}{\eta N_{\rm eff}} \right]}  \pm \frac{\delta_{\Delta}}{\ln\left[\frac{N_{\rm eff} + 1}{\eta N_{\rm eff}} \right]}\non\\
\implies \frac{\delta T_{\Delta}}{T} &=& \frac{\delta_{N_{\rm eff}}}{N_{\rm eff}\left(N_{\rm eff} + 1 \right)\ln\left[\frac{N_{\rm eff} + 1}{\eta N_{\rm eff}} \right]} + \frac{\delta_{\Delta}}{T\ln \left[\frac{N_{\rm eff} + 1}{\eta N_{\rm eff}} \right]} \approx \frac{\delta_{N_{\rm eff}}}{N_{\rm eff}\left(N_{\rm eff} + 1 \right)\ln\left[\frac{N_{\rm eff} + 1}{\eta N_{\rm eff}} \right]} + \frac{\delta_{\Delta}}{\omega_{-1}}.
\ea
\end{widetext}
Here we have used  Eqs. \eqref{etapg} and \eqref{kap}, and neglected factors of the order $\delta_{\rm N_{eff}}^2$ and  $\delta_{\rm N_{eff}} \delta_{\Delta}$. One 
can reduce the uncertainty $\delta_{N_{\rm eff}}$ in the estimation of $N_{\rm eff}$ by increasing the number of measurements, and neglect the uncertainty caused by $\delta_{\Delta}$ 
for $\delta_{\Delta} \ll \omega_{-1}$.

\section{Optimal thermometry}
\label{appC}

The QFI Eq. \eqref{Hal} is a monotonically decreasing function of $\eta$. The maximal (optimal) low-temperature QFI is obtained from Eq. \eqref{Hal} for $\eta = 1$ (for a given 
$\omega_{-1}$ and $T$), 
which according to Eq. \eqref{etapg} requires that the bath 
coupling spectrum weighted by the Floquet harmonic probability satisfies
\ba
P_{1}G(\omega_{-1}) \gg P_{0}G(\omega_{0}) , P_{1}G(\omega_{1}). 
\label{pmgm}
\ea
Under this condition we obtain
\ba
\mathcal{H} = \mathcal{H}_{\rm max} \approx \frac{e^{\frac{\omega_{-1}}{T}} \omega_{-1}^2}{\left(-1+ e^{\frac{\omega_{-1}}{T}}\right)^2 T^4}.
\label{Hopt}
\ea
Condition \eqref{pmgm} can be satisfied by Ohmic spectra with low cutoff frequency.
By contrast,  the corresponding QFI in the absence of control,
\ba
\mathcal{H} \approx \frac{e^{\frac{\omega_{0}}{T}} \omega_0^2}{\left(-1+e^{\frac{\omega_{0}}{T}}\right)^2 T^4} \approx \frac{\omega_0^2}{T^4}e^{-\omega_0/T} \to 0,
\label{H0nocont}
\ea
is significantly smaller than that obtained in Eq. \eqref{Hopt} for temperatures $T\sim \omega_{-1} \ll \omega_0, \omega_{1}$, thus revealing the advantage of our proposed control scheme.

\section{Thermometry for a nearly flat bath spectrum}
\label{appD}

\begin{widetext}
\begin{figure*}[ht]
\begin{center}
\includegraphics[width= 1.7\columnwidth, angle = 0]{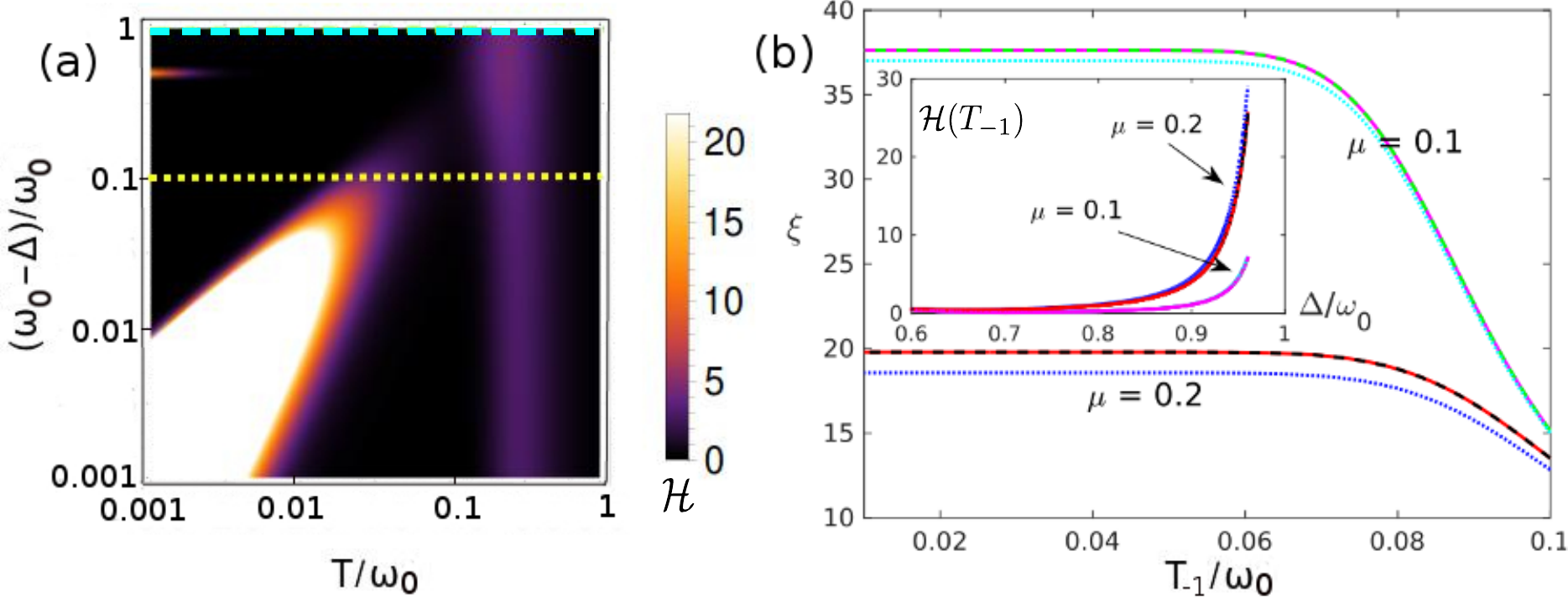}
\end{center}
\caption {(a) {\bf Quantum Fisher information:} Quantum Fisher information (QFI) $\mathcal{H}$ as a function of $\left(\omega_0 - \Delta\right)/\omega_0$ and $T/\omega_0$ (in log-log scale) for sinusoidal modulation (Cf. Eq. \eqref{omtsup}) 
with a harmonic oscillator DCQT,  probing a bath with a nearly flat bath spectrum. Here we have  considered the zeroth and first three sidebands ($m = 0,\pm 1, \pm 2, \pm 3$). As seen from the plot, 
$\mathcal{H}$ is large as $T \to 0$ and $\Delta \uparrow \omega_0$. The turquoise dashed line at $\Delta = 0$ corresponds to thermometry in absence of control, and the 
yellow dotted line shows the variation of QFI with 
temperature for 
$\Delta = 0.9$ (see Fig. 2 of main text). The second sideband results in non-zero QFI for small temperatures close to $2\Delta \uparrow \omega_0$. 
The color scheme for $\mathcal{H}$ is shown in the colorbar CS.
(b) {\bf Relative error bound:} The relative error bound $\xi$ per measurement ($\mathcal{M} = 1$), for estimation of bath temperature  at $T = T_{-1} = (\omega_0 - \Delta)/4$ remains finite and 
approximately constant, even as $T_{-1}$ approaches absolute zero (Cf. Eq. \eqref{xisup}). The blue (cyan) dotted line, red (magenta) dashed line and black (green) solid line 
show $\xi$ for $\mu = 0.2$ ($\mu = 0.1$) considering only the sidebands $m = 0, \pm 1$, $m = 0,\pm 1, \pm 2$ and $m = 0,\pm 1,\pm 2, \pm 3$ respectively. As seen above,
higher sidebands have negligible effect on $\xi$, specially for 
small $\mu$. The results with $m = 0,\pm 1, \pm 2$ and $m = 0,\pm 1,\pm 2, \pm 3$ overlap, showing that inclusion of the higher sidebands fails to change the results in any significant way.
Inset:  QFI $\mathcal{H}$ at $T = T_{-1}$ increases with increasing $\Delta$, finally diverging as $\Delta \uparrow \omega_0$, for $\mu = 0.1$ and $\mu = 0.2$. QFI does not
change significantly for analysis
with sidebands $m = 0, \pm 1$; $m = 0,\pm 1, \pm 2$ and $m = 0,\pm 1,\pm 2, \pm 3$.
Here $\omega_0 = 1$, and we have considered a nearly flat bath spectral response (Eq. \ref{nfbs}) with $G_0 \ll \omega_{\rm min} \ll T_{-1}$.}
\label{figm1}
\end{figure*}
\end{widetext}

The significant improvement in temperature estimation offered by dynamical control is not restricted to bath spectra satisfying the condition Eq. \eqref{pmgm}. Namely, in what
follows, we consider
a nearly flat bath spectral response, defined by 
\ba
G(\omega \geq \omega_{\rm min} > 0) \approx G_0 > 0,~~G(\omega \to 0) \to 0.
\label{nfbs}
\ea
Under the above conditions, Eq. \eqref{neffgenho} simplifies to
\ba
N_{\rm eff} &=& \frac{1}{e^{\omega_{\rm eff}/T} - 1},\\
e^{-\omega_{\rm eff}/T} &=& (1-\frac{\mu^2}{2})e^{-\omega_0/T} \non\\&+& \frac{\mu^2}{4}\left(e^{-\omega_{-1}/T} + e^{-\omega_{1}/T}\right)\non,
\label{neffnfbsho}
\ea
where we have taken 
\ba
P_0 \approx 1-\frac{\mu^2}{2} \quad \quad \text{and} \quad \quad P_{\pm 1} \approx \frac{\mu^2}{4}.
\ea
One can use Eqs. \eqref{qfignsn} and \eqref{neffnfbsho} to evaluate the QFI for sinusoidal modulation in a simple harmonic oscillator thermometer:
\ba
\mathcal{H} \approx \frac{A_{\sin}}{B_{\sin}},
\ea
where (see Fig. \ref{figm1}),
\begin{widetext}
\ba
A_{\sin} &=& 16 e^{\frac{3 \Delta + \omega_0}{T}} \Big[\left(-2+\mu ^2\right) \omega_0+\mu ^2 \Big(-\omega_0
\cosh\left(\frac{\Delta}{T}\right) +\Delta  \sinh\left(\frac{\Delta }{T}\right)\Big)\Big]^2,\non\\
B_{\sin} &=& T^4 \left(\mu ^2+e^{\frac{\Delta}{T}} \left(4-4 e^{\frac{\omega_0}{T}}-2 \mu ^2+e^{\frac{\Delta }{T}} \mu ^2\right)\right)^2 \left(\mu ^2+e^{\frac{\Delta }{T}}
\left(4+\left(-2+e^{\frac{\Delta }{T}}\right) \mu ^2\right)\right).
\label{qfisin}
\ea
\end{widetext}

QFI under sinusoidal control, Eq. \eqref{qfisin}, given by
can be rewritten as the sum of three terms that can be ascribed to the $m = -1, 0, 1$ sidebands:
\begin{widetext}
\ba
\mathcal{H}&\approx& \mathcal{H}_{-1} + \mathcal{H}_{0}+\mathcal{H}_{1}, \non\\
\mathcal{H}_{\pm 1} &=& \frac{P_{\pm 1} e^{\frac{\omega_{\pm 1}}{T}} \omega_{\pm 1}^2}{\left(-1+e^{\frac{\omega_{\pm 1}}{T}}\right)^2 T^4};~~ \mathcal{H}_{0} = \tilde{\mathcal{H}}_{0} + \mathcal{H}_{\rm rem}\non\\
\tilde{\mathcal{H}}_{0} &=& \frac{P_0 e^{\frac{\omega_0}{T}} \omega_0^2}{\left(-1+e^{\frac{\omega_0}{T}}\right)^2 T^4}, \non\\
\mathcal{H}_{\rm rem} &=& \mathcal{H} - \left(\mathcal{H}_{-1} + \tilde{\mathcal{H}}_{0} + \mathcal{H}_{1} \right) = \frac{A_{\sin}}{B_{\sin}} - \left(\mathcal{H}_{-1} + \tilde{\mathcal{H}}_{0}+ \mathcal{H}_{1}\right), \non\\
 &=& \frac{1}{4 T^4}e^{\omega_0/T} \Big[-\frac{e^{\Delta /T} \mu ^2 (\Delta -\omega_0)^2}{\left(e^{\Delta /T}-e^{\omega_0/T}\right)^2} 
 +\frac{2\left(-2+\mu ^2\right) \omega_0^2}{\left(-1+e^{\omega_0/T}\right)^2} -\frac{e^{\Delta /T} \mu ^2 (\Delta +\omega_0)^2}{\left(-1+e^{\frac{\Delta +\omega_0}{T}}\right)^2} + \frac{l_1}{l_2}\Big], 
\label{chiapp}
\ea
\end{widetext}
where
\begin{widetext}
\ba
l_1 &=& 64 e^{\frac{3\Delta}{T}} \Big(\left(-2+\mu^2\right)\omega_0 +\mu^2 \left(-\omega_0 \cosh\left[\frac{\Delta}{T}\right]+\Delta  \sinh\left[\frac{\Delta}{T}\right]\right)\Big)^2, \non\\
l_2 &=& \left(\mu^2+e^{\Delta /T} \left(4-4 e^{\omega_0/T}-2 \mu^2+e^{\Delta /T} \mu^2\right)\right)^2 \Big(\mu^2 + e^{\Delta /T} \left(4+\left(-2+e^{\Delta /T}\right) \mu^2\right)\Big).
\ea
\end{widetext}
In the limit of low temperatures, $\mathcal{H} \approx \mathcal{H}_{-1} \gg \mathcal{H}_{0}, \mathcal{H}_{1}$. One can show that in this regime, assuming
$T\to 0$, both $\mathcal{H}_{-1}$, and $\mathcal{H}$, 
attain maximum at $T \approx T_{-1} = \omega_{-1}/4$. Upon 
replacing $T$ by $T_{-1} \to 0$ in the expression for $\mathcal{H}_{-1}$ in Eq. \eqref{chiapp}, one gets
\ba
\lim_{\Delta \uparrow \omega_0}\xi\left(T = T_{-1} \to 0\right) &\approx& \frac{1}{T_{-1}\sqrt{\mathcal{H}_{-1}} \mathcal{M}} \non\\&\to& \frac{e^2}{2\mu\sqrt{\mathcal{M}}}.
\label{xisup}
\ea

One can estimate the range of $\mu, \Delta$ for which the first sideband dominates at $T \approx T_{-1}$, thereby giving rise to a maxima there, by defining
the quantity
\ba
R = \ln \frac{\mathcal{H}_{-1}(T_{-1})}{{\rm Abs}\left[\mathcal{H}(T_{-1}) - \mathcal{H}_{-1}(T_{-1})\right]}.
\label{appR}
\ea
A positive $R$ denotes $\mathcal{H}_{-1}(T = T_{-1})$ is the major contribution to $\mathcal{H}(T = T_{-1})$, thus resulting in a maxima at $T \approx T_{-1}$.

 For $\mu$ that is small, but finite, and $2\Delta \uparrow \omega_0$, the second sideband enhances the QFI marginally
for $T \sim \omega_0 - 2\Delta$ (Cf. Fig. \ref{figm1}). However, as mentioned before, the contributions of the higher sidebands decrease rapidly for small $\mu$. 
Optimal control of low-temperature thermometry would demand $\Delta \uparrow \omega_0$, in which case only the first sidebands 
contribute ($m = 0, \pm 1$) significantly.

\section{Sub-Ohmic bath spectrum}
\label{appE}

Here we focus on  bath spectral-response functions satisfying Eq.~(4) of the main text
and the KMS condition Eq. \eqref{kmsN}, in the $s < 1$ regime that corresponds to sub-Ohmic bath spectrum.
The QFI and relative error bound $\xi$ are independent of $\gamma$,  where we consider $\gamma > 0$ to be  small enough to justify the secular approximation
(see App. \ref{appA}). In 
the limit of small $\gamma$, $s \to 0$ and $\omega_c \to \infty$, the solution of the equation 
\ba
\frac{\partial G(\omega)}{\partial \omega} = 0
\label{flatG}
\ea
implies that one gets a nearly flat bath spectrum $G(\omega) \approx G_0 \approx \gamma s^s \omega_c \exp (-s)$
for
$\omega_{\rm min} \lesssim \omega \ll \omega_c$, with
$\omega_{\rm min}  \approx s \omega_c$. Even though $s \to 0$, we have 
\ba
\lim_{s\to0} s^s &=& \lim_{s \to 0} e^{s\ln s} = \exp\left[\lim_{s \to 0}\frac{\ln s}{1/s}\right] =  \exp\left[-\lim_{s\to 0}\frac{1/s}{1/s^2}\right] \non\\&=& \lim_{s\to0} e^{-s} = 1,
\label{limits}
\ea
which can eventually result in a finite $G_0$. In deriving Eq. \eqref{limits} we have used L'Hospital's rule. Here the secular approximation demands 
$G_0 \ll \omega_{\rm min}$.

Thermometry with DCQT in this NFBS limit of sub-Ohmic bath spectrum would allow us
to estimate temperatures with a constant and small relative 
error very close to the absolute zero. It would also enable us to measure low temperatures with high precision for a broad variety of bath spectra, as long as 
$P_1 G(\omega_{-1} \sim T)$ is large enough to contribute significantly to the dynamics, and consequently to the QFI.

\section{QFI with more than two peaks}
\label{appF}

Depending on the form of QFI we wish to engineer, one can extend the control scheme detailed in the main text to arbitrary periodic
modulations of $\omega(t)$. For example, one can adapt the control scheme to the generation of
 multiple (two or more) peaks of QFI, which would in turn allow precise thermometry in the vicinity of all the temperature values which 
 correspond to  QFI peaks. As before, below we consider a 
DCQT aimed at measuring  a temperature $T$ of a thermal bath with a NFBS.
Similar arguments apply for probing  a multimode bath, 
with different modes $k$ thermalized at different temperatures $T_k$. In either case, it is advantageous to choose a modulation of the form (with $m^{\prime} = l, l^{\prime}$)
\ba
\omega(t) = \omega_0 + \Delta\left(\mu_{l}\sin\left(lt \Delta\right) + \mu_{l^{\prime} \neq l}\sin\left(l^{\prime}t \Delta\right)\right),
\label{eq3peak}
\ea
which gives rise to  significant $P_m$'s only for the sidebands close to $m = 0, l,l^{\prime}$. 

Proper tuning of the parameters $l,\mu_l, l^{\prime},  \mu_{l^{\prime}}$, and $\Delta$  can give rise to 
QFI with 3 peaks of our choice, allowing us to accurately measure a wide range of temperatures (Cf. Fig. 4 of main text). 
Similarly, one can increase the number of frequency components 
in $\omega(t)$ in order to increase the number of peaks in the QFI, thus broadening the range of temperatures that can be measured accurately with our DCQT.  We note 
that a recent work has
studied the emergence of multiple peaks of QFI by considering a probe with multi-gapped spectra, and highly degenerate excited states \cite{campbell18precision}.
In contrast, our control scheme enables us to achieve
multi-peak probing by a non-degenerate system characterized by a single energy spacing, yet with a QFI tunable
according to our temperature(s) of interest.

\bibliographystyle{naturemag}

\end{document}